\def\aj{AJ}                   
\def\araa{ARA\&A}             
\def\apj{ApJ}                 
\def\apjl{ApJ}                
\def\apjs{ApJS}               
\def\apss{Ap\&SS}             
\def\aap{A\&A}                
\def\aaps{A\&AS}              
\def\ba{Baltic Astr.}         
\def\mnras{MNRAS}             
\def\pasj{PASJ}               
\def\sci{Science}             
\def\ssr{Space~Sci.~Rev.}     
\def\nat{Nature}              

\def\gca{Geochim.~Cosmochim.~Acta}   
\def\jqsrt{J.~Quant.~Spec.~Radiat.~Transf.}

\def\physscr{Phys.~Scr}   

\newcommand{\Msun}{M$_\odot$}

\documentclass[]{svmult}
\usepackage{epsfig,psfig}

\begin{document}

\title{Atomic and Molecular Data for Stellar Physics: Former Successes
  and Future Challenges}
\author{
  A. Jorissen\thanks{Research Associate FNRS, Belgium}}
\institute{
Institut d'Astronomie et d'Astrophysique,
  Universit\'e Libre de Bruxelles, Belgium.}

\maketitle

\section*{Abstract}

This review highlights current (and future!) hot topics in
astrophysics where atomic or molecular input data are (or will be)
essential,
with special emphasis on topics relating to
nucleosynthesis and cosmochemistry.

We first discuss issues (like the abundances of oxygen and iron in
the Sun, and that of lithium in post-AGB stars) 
where the use of poor-quality atomic or molecular data have led to
spurious astrophysical puzzles which sparked fancy
astrophysical models or theories.
We then address issues where the advent of new instruments (like the
ultraviolet high-resolution spectrographs---GHRS onboard
HST, Keck-HRS or VLT-UVES---or future infrared satellites)
calls for new and accurate atomic or molecular data.
\medskip\\
{\bf Keywords.} atomic physics -- molecules -- oscillator strengths --
r-process -- s-process -- cosmochronology -- stellar atmospheres --
nucleosynthesis.

\clearpage

\twocolumn

\section{Introduction}

Despite many reviews highlighting 
the interplay between astrophysics and atomic/molecular
physics
\cite{Gustafsson-95,Gustafsson98,Kurucz02a,Kurucz02,Lambert-94}, 
the importance of using
as good microscopic data as possible is often overlooked. As perfectly 
expressed by Kurucz \cite{Kurucz02},  
\textsl{astrophysicists work on ``Important'', ``Big'' problems and
  think that the basic physics that they require to solve their
  problems has already been done, or, if it has not been done, it is
  easy and can be readily produced, as opposed to the hard problems
  they are working on. They have it backward. Getting the basic data
  is the hard part (...).
\smallskip\\
\indent Half the lines in the solar spectrum are not identified. All the
features are blended. Most features have unidentified components that
make it difficult to treat any of the identified components in the
blend. And even the known lines have hyperfine and isotopic splittings 
that have not yet been measured. Is an asymmetry produced by a
splitting, or by a velocity field, or both? It is very difficult to
determine abundances, or any property, reliably, when you do not know
what you are working with.
}

The present review is a compendium of astrophysical questions where atomic or molecular
data have played (or will play in the near future) a central role. Cases where the use of
incorrect or incomplete atomic data led to the development of spurious astrophysical models or
theories will be especially stressed. Quotations from the original papers pleading for more or
better atomic or molecular data have been systematically included in
this text in \textsl{slanted face}. Reflecting the
author's own interests,  this review is biased towards topics relating to nucleosynthesis,
stellar atmospheres, and cosmochemistry. Only the ultraviolet, visible and
infrared spectral windows will be touched upon here. For a discussion of the needs in atomic and
molecular data in other spectral windows (like X,
$\gamma$ or radio) and for issues relating to extragalactic astronomy, cosmology or planetary
atmospheres, the reader is referred to the invited reports from the {\it Laboratory Space
Science Workshop} (April 1998; to be found on  {\tt
http://cfa-www.harvard.edu/ amdata/ampdata/law/reports.html}). Other conferences of interest 
include {\it Molecules in the Stellar Environment} (1994, IAU
Coll.~146; \cite{IAUC146}), {\it Astrophysical Applications of Powerful
New Databases} (1995;
\cite{Databases-95}), {\it Laboratory and Astronomical High Resolution
Spectra} (1995;
\cite{Sauval}), {\it Atomic and Molecular Data for Astrophysics: New
Developments, Case Studies and Future Needs} (2000; \cite{IAU24JD1}), the
report of the Leiden meeting on the {\it Preparatory work for the
Herschel Space Observatory} (October 2001; 
{\tt http://www.astro.rug.nl/ $\sim$european/}), {\it Stellar Atmospheric
Modeling} (September 2001; \cite{ASP288}), the {\it NASA Laboratory
Astrophysics Workshop}   (2002; {\tt
http://www.astrochemistry.org/ nasalaw.html} or  NASA Conference
Proceeding 2002-211863), and finally {\it Modeling of Stellar
Atmospheres} (2003, IAU Symp. 210; 
\cite{IAUS210}).

This review is organized as follows. 
Section~\ref{Sect:nucleosynthesis}
sets the stage  with
a few basic facts about nucleosynthesis. 
Section~\ref{Sect:hot} then illustrates 
the cross-fertilization between atomic physics and
astrophysics on particular examples. First, we present several astrophysical
puzzles which were 
solved once that more accurate or more complete atomic data became
available
(the iron and oxygen abundances in the Sun, the lithium abundance in
post-AGB stars, the galactic age based on the Th/Nd, Th/Eu and Th/U
chronometers,  
the excitation mechanism of the pulsation in $\beta$ Cephei stars, and the
meteorology of brown dwarf stars).   Second, we discuss issues where
the advent of new instruments calls for new atomic or molecular data
(the exploration of the 300--400~nm ultraviolet window with high-resolution
spectrographs, and of the near- and far-infrared windows with the
{\it ISO} and {\it Herschel} satellites). In Section~\ref{Sect:success}, 
we list recent achievements of the atomic-physics community of special 
importance for astrophysicists (new oscillator strengths: solar,
measured or computed {\it ab initio}; better partition functions; 
new theory for Van der Waals
broadening by neutral hydrogen atoms: the {\it damping enhancement
  factor} is gone; 3-D modeling of atmospheric
convection: {\it microturbulence} is gone). Finally,
Section~\ref{Sect:databases} provides starting points to locate
existing databases of atomic and molecular data.    

\section{The solar-system abundance curve: A clue to nucleosynthesis}
\label{Sect:nucleosynthesis}

\begin{figure}
\centerline{\psfig{file=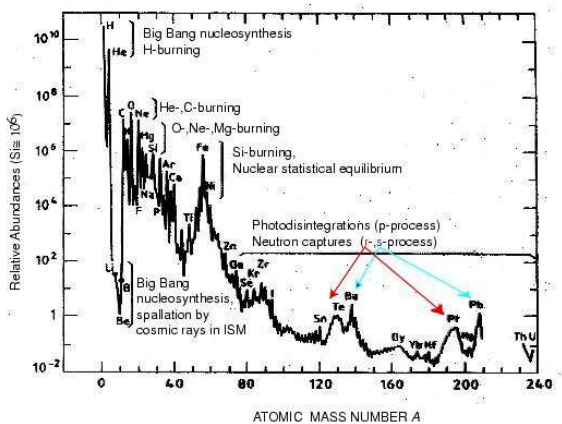,width=8.5cm}}
\caption{\label{Fig:abundances}
Abundances of the chemical elements in the solar system
\cite{Anders-Grevesse-89}, and the major nucleosynthesis processes
responsible for their production. 
}
\end{figure}

The relative abundances of the chemical elements in
the solar system have been  derived for the first time by Russell in 1929 
\cite{Russell29}, from the spectroscopic analysis of the solar photosphere,  
and by Goldschmidt in 1937 \cite{Goldschmidt37}, 
from the chemical analysis of primitive 
meteorites known as  carbonaceous chondrites. More recent compilations and 
revisions of the solar-system abundances were provided by 
Suess \& Urey in 1956 \cite{Suess-Urey-57}, 
by Anders \& Grevesse in 1989 \cite{Anders-Grevesse-89} and by Grevesse \& Sauval in 1998 \cite{Grevesse-1998}
(Fig.~\ref{Fig:abundances}).
The solar-system abundances hold the key to the various nucleosynthesis processes
at work in the  universe. In a seminal paper (that has become known as B2FH \cite{B2FH}),
Burbidge, Burbidge, Fowler \& Hoyle proposed in 1957 a small number of distinct nucleosynthesis
processes   accounting for the major features of the abundance curve. Although nearly fifty
years have elapsed, their
analysis remains basically correct, as recently reviewed by \cite{Wallerstein-97}.

\begin{figure}
\centerline{\psfig{file=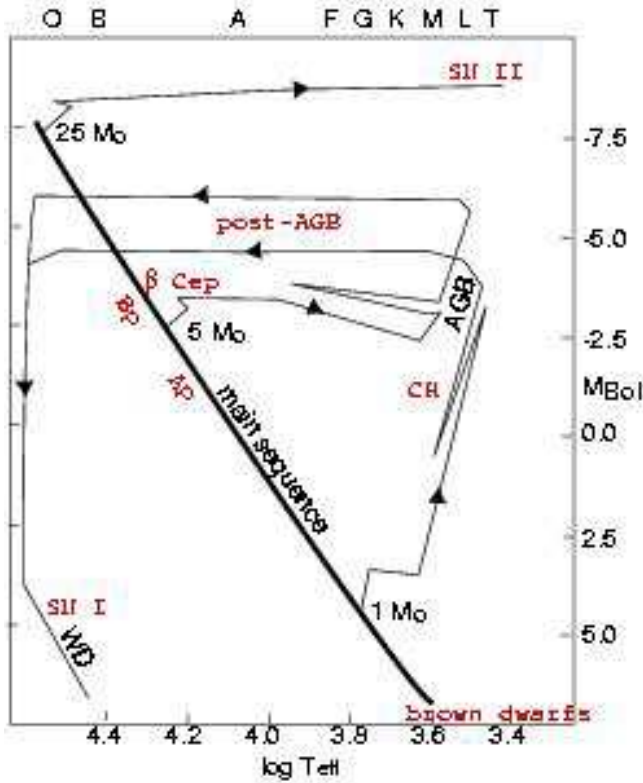,width=8.5cm}}
\caption{\label{Fig:HR}
Hertzsprung-Russell (HR) diagram showing the location of the various
classes discussed in the present paper:  chemically-peculiar main sequence Ap and Bp stars, $\beta$ Cephei
variables, AGB stars, post-AGB stars, lead (CH) stars,
type~Ia supernovae, type~II supernovae, low-mass stars (T and L brown dwarfs). 
}
\end{figure}

The various nucleosynthesis processes at work in the universe are the
following \cite{B2FH,Wallerstein-97}:\\
(i) {\it cosmological nucleosynthesis}. At the time of the Big Bang, only the 
lightest elements were formed, from hydrogen to lithium; \\
(ii) {\it spallation reactions} in the interstellar medium are still
producing lithium, 
beryllium and boron nowadays;\\
(iii) {\it the Cameron-Fowler mechanism} (\cite{Cameron-Fowler-71} and Sect.~\ref{Sect:Li})
produces lithium under very special circumstances in warm convective zones of late-type giant
stars;\\
(iv) {\it the major nuclear-burning stages responsible for the energy generation in
stars}  produce helium, as well as the elements from carbon to iron;\\
(v) {\it the nuclear statistical equilibrium}, established in 
supernovae of types~I and II, is responsible for the build-up of the
so-called iron peak, extending roughly from
Ti to Zn and involving the elements with the highest binding energies;\\
(vi) {\it the s-, r- and p-processes} are responsible for the synthesis of nuclides 
heavier than $^{56}$Fe, the so-called ``heavy elements'' ($Z > 30$). 
The p-process produces isotopes on the proton-rich
side of the valley of $\beta$ stability. It operates through a chain of photodissociation
reactions in the hot environment of some type~II supernova explosions and
will not be discussed here.  The s- and r-processes are neutron-capture processes producing
isotopes within the valley  of
$\beta$ stability or on its neutron-rich side. The s-process (where {\it s} stands
for {\it slow}) operates in low neutron-density environments associated with helium-burning in
asymptotic giant branch (AGB) stars (Fig.~\ref{Fig:HR}; central
He-burning in massive stars also contributes to the production of
s-process elements between Fe and Sr). The s-process imprints a very specific
signature on the abundance curve in the form of peaks around the so-called ``magic nuclei''
(Fig.~\ref{Fig:abundances}), which have a closed neutron shell (the analogues of the rare gases
which have a closed electronic shell in atomic physics). The first
peak involves 
$^{88}$Sr, $^{89}$Y, $^{90}$Zr, the second involves $^{138}$Ba,
$^{139}$La and $^{140}$Ce whereas  the
third peak involves $^{208}$Pb and $^{209}$Bi. 

The r-process (where {\it r} stands
for {\it rapid}) operates in high
neutron-density environments, probably associated with type~II supernova explosions or 
with the coalescence of two neutron stars. 
It produces abundance peaks which resemble those of the s-process, but
which are shifted to the left by about 15
mass units (Fig.~\ref{Fig:abundances}). Elements with a large
contribution from the  r-process include  Eu, Gd, W, Re, Ir, Pt, as
well as 
the actinides Th and U.


Astrophysical diagnostics of some among these processes will be discussed in
Sect.~\ref{Sect:hot}, 
with special emphasis on the atomic-physics input. 
They involve stars at different evolutionary stages, 
all displayed on Fig.~\ref{Fig:HR}.
 
\section{Astrophysical puzzles and atomic physics clues}
\label{Sect:hot}

Atomic and molecular data are used in astrophysics for two different purposes:
(i) to compute the opacity of the stellar matter,
and hence, to derive the thermal structure of the star and its
atmosphere (i.e., the run
of $T$, $P$ and $\rho$ as a function of distance from the center; see
\cite{Gustafsson-95} and the contributions by J\o rgensen, Liebert and Scholz \& Wehrse in 
\cite{IAUC146} for a
discussion of this specific aspect), or (ii) to predict abundances for specific chemical
elements from the equivalent widths of appropriate spectral lines or from spectral synthesis.

Abundances are generally derived under the assumption that matter and radiation are locally
in equilibrium with each other (the so-called {\it Local Thermodynamical 
Equilibrium} hypothesis, denoted LTE). This hypothesis
does not  hold for specific lines under specific  stellar conditions (see
\cite{Lambert-94} for a detailed review).  If abundances must be derived under non-LTE conditions,
specific inputs will be  required from atomic physics. 
They will not be discussed here,  however.

\subsection{The iron problem}
\label{Sect:Fe}

\begin{figure}
\centerline{\psfig{file=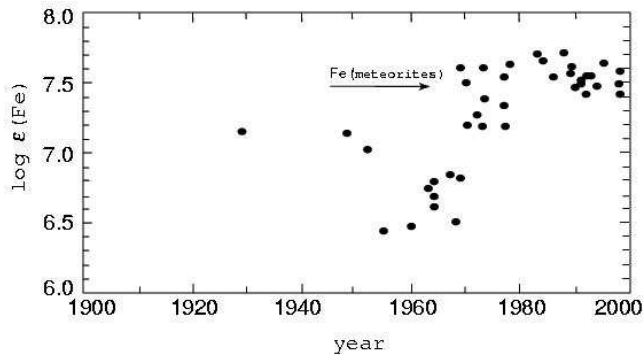,width=8.5cm}}
\caption{\label{Fig:Fe}
The evolution of the solar iron abundance (in the scale where $\log
\epsilon(\mathrm{H}) = 12.$) with time over the
last century.
From \cite{Grevesse99}.
}
\end{figure}

Iron is the prototypical element used to estimate the so-called
stellar metallicity [Fe/H] = $\log (\epsilon({\rm Fe})_*) - \log
(\epsilon({\rm Fe})_\odot)$, where $\log (\epsilon({\rm Fe}))$ denotes the Fe
abundance in the logarithmic scale where $\log (\epsilon({\rm H})) =
12$, and subscript $\odot$ 
refers to the solar abundance. 
The iron abundance is used as a normalization\footnote{The logarithmic
  normalized abundance of element X is generally denoted [X/Fe], where 
  [X/Fe] $= \log (\epsilon({\rm X})/\epsilon({\rm Fe}))_* - \log (\epsilon({\rm
    X})/\epsilon({\rm Fe}))_\odot$} 
element in most stellar
abundance works---because it has a wealth of lines in the optical
spectra of cool stars---hence its central
importance in cosmochemistry. And yet, even in the Sun, the situation
concerning the Fe abundance was very unsatisfactory until just a few
years ago,  the abundances derived from Fe~I lines being
often larger than the Fe~II abundance by as much as 0.15~dex. 
Although the solar Fe~II abundance was in agreement with the
meteoritic value ($\log(\epsilon({\rm Fe})) = 7.50$; \cite{Grevesse-1998} and
Fig.~\ref{Fig:Fe}),  it had uncomfortably large uncertainties
approaching 25\% \cite{Grevesse99}.    

The history of the determination of the solar iron abundance  over the last
century is indeed very shaky, with atomic physics playing a
central role, as reviewed by \cite{Grevesse99}.
Fig.~\ref{Fig:Fe} shows the evolution of the iron abundances derived
over the years starting from its very first measurement by Russell in 1929
\cite{Russell29}. Early determinations resulted in abundances an order of
magnitude below the coronal and meteoritic values. This discrepancy 
led to a flurry of new astrophysical models aiming at explaining the
origin of the iron fractionation between the photosphere and the
corona... until it was discovered in 1969 that this fractionation is
in fact spurious and resulted from a systematic error in the absolute scale of
the transition probabilities \cite{Garz69}!    

The following milestone was the publication of measured transition
probabilities
(\cite{Biemont-91,Blackwell84,Fuhr-88,Holweger91}  and
references therein) which resulted in a spectacular reduction of the
scatter in the early 1980s. Still, in the 1990s, a debate raged about
which from the low Fe abundance ($\log(\epsilon({\rm Fe})) = 7.50$) or
high Fe abundance ($\log(\epsilon({\rm Fe})) = 7.67$)
is the correct one (see also the review in \cite{Lambert-96}). The former value, advocated
by the Kiel-Hannover group \cite{Holweger91,Holweger95},  
agrees with the meteoritic value \cite{Grevesse-1998}, whereas the latter was defended by the
Oxford  group \cite{Blackwell84,Blackwell95a,Blackwell95b}.
The origin of this longstanding puzzle has been
identified by \cite{Grevesse99} as the cumulative effect of slight
differences between the equivalent widths, the $f$-values\footnote{In the remainder of this paper, $f$ denotes
oscillator strengths} absolute scales, the microturbulent velocities and the empirical enhancement
factors of the damping constants (see Sect.~\ref{Sect:OMara}) 
adopted by the two groups (see also \cite{Asplund-00a}). The photospheric Fe abundance 
($\log(\epsilon({\rm Fe})) = 7.50$) derived by \cite{Grevesse99} is now
in perfect agreement with the meteoritic value.

Despite the satisfactory situation regarding the solar Fe abundance, 
it must be stressed that the number of Fe (especially Fe~II) lines
with experimental $f$ values remains uncomfortably small.  
In their evaluation of the solar Fe abundance based on the currently 
best set of atomic data
and lines, Grevesse \& Sauval \cite{Grevesse99} use 65 Fe~I lines but
only 13 Fe~II lines with experimental $f$ values 
(see also \cite{Fuhr-88,Lambert-96}). 
For Fe~II, the situation is especially worrisome since the dispersion of
the abundances derived from the various individual Fe~II lines 
remains quite large (0.095~dex; Fig.~7 of \cite{Grevesse99}). 
The situation
may be expected to improve rapidly, though, thanks to the FERRUM
project (\cite{Johansson02a,Johansson02b}; see Sect.~\ref{Sect:loggf}).

\subsection{The oxygen problem}
\label{Sect:O}

Oxygen is a very important element for nucleosynthesis: the run 
of the O/Fe ratio  with
metallicity [Fe/H] (Sect.~\ref{Sect:Fe}) provides
interesting clues as to their respective production sites, and the
lifetimes of these sites in the Galaxy. Type~II supernovae (originating from massive, short-living stellar
progenitors) are responsible for the production of oxygen at the galactic level, whereas Types~I (originating
from low-mass, long-living stellar progenitors) and II  contribute to Fe \cite{Wallerstein-97}.
The iron production has thus started at a later stage of galactic chemical
evolution than the oxygen one.
Hence [O/Fe] should increase with decreasing [Fe/H]. But the exact shape of the [O/Fe] vs.
[Fe/H] relation is currently very polemical
\cite{Allende01}. A first group of astronomers defends a monotonic increase of [O/Fe] with decreasing [Fe/H],
whereas a second group  prefers an essentially constant [O/Fe] for old stars with [Fe/H] $\le -1$.

The solar-system oxygen abundance is not free from debate either. 
Since oxygen, being too volatile, is not present in its elemental form in
meteorites, its solar-system abundance must
rely on spectroscopic determinations in the solar photosphere
or corona. There are four different spectroscopic diagnostics
available to infer the oxygen abundance in the Sun
and in cool stars \cite{Allende01,Asplund03,Bessell02}: 
(i) permitted O~I lines at $\lambda$ 777~nm
(the oxygen triplet), 844.67, 926.26 and 926.59~nm;
(ii) the  $\lambda 630.0321$ and $\lambda 636.3792$ [O I] doublet;
(iii) electronic $A\Sigma - X\Pi$ UV OH lines;
(iv) vibration-rotation or pure rotation  OH lines in the infrared.

For a long-time, spectroscopists faced, however, the so-called oxygen problem
\cite{Allende01,Bessell02}: 
the four diagnostics did not yield consistent abundances. 
Moreover, several  determinations of the solar oxygen abundance  gave values significantly
higher (by up to 0.2~dex) than the  oxygen abundances in both the local interstellar medium and
in the photospheres of hot stars in the solar neighborhood\footnote{A
  similar discrepancy, not yet resolved, exists for the solar-system
  abundance of  fluorine, which is about 0.1~dex lower than the value
  observed in nearby solar-metallicity stars
  \cite{Cunha-03,Jorissen-92}} (\cite{Allende01}
and references therein), which typically fall in the range
$\log(\epsilon({\rm O})) = 8.65$--8.70.  

Which diagnostic should one use to determine the oxygen abundance in cool
stars and how does one 
reconcile the different results from the different diagnostics? 

It is now clear that the solution to the ``oxygen
problem'' must no longer be looked for 
on the atomic and molecular physics side, which has made 
substantial progress in recent years (as reviewed by
\cite{Asplund03}). It has been shown, for example, that: (i)
the forbidden [O~I] $\lambda 630.0321$ line is blended with Ni~I
$\lambda 630.0363$ \cite{Lambert78}, whose $f$-value has recently been measured
\cite{Johansson03}, and which represents about 50\% of the blend in
the Sun \cite{Asplund03};\\
(ii)  the forbidden [O~I] $\lambda 636.3792$ line is blended with CN
lines \cite{Asplund03,Lambert78} having poorly determined parameters;\\
(iii) three fine-structure components have recently been discovered
for each of the permitted O~I $\lambda$ 926.26 and 926.59
lines \cite{Asplund03}. Neglecting these fine-structure components
would increase the estimated abundance by 0.1~dex! 

The almost satisfactory situation reached on the atomic and molecular physics
side now makes it clear  that the oxygen problem has its roots in  unaccounted non-LTE effects and in the
1-D rather than 3-D description  of convective stellar atmospheres. 
Using a fully 3-D model of the solar atmosphere, 
Asplund et al. \cite{Asplund03}
(see also Sect.~\ref{Sect:3D}) have settled the oxygen problem for the Sun
(with an abundance of $8.74\pm0.06$, after correcting for non-LTE effects), 
and it may be
hoped that a similar treatment applied to cool stars will remove
the existing discrepancies \cite{Bessell02}.

Allende Prieto et al. \cite{Allende01} conclude their review of the oxygen problem by
stating that it \textsl{exemplifies the
importance of detailed
  line profiles and accurate wavelength scales in chemical analyses of 
  stars from spectra as well as the need for hydrodynamical model
  atmospheres in fine analyses of stellar spectra}. More on the second 
plea will come in Sect.~\ref{Sect:3D}.

\subsection{Cerium: The lithium substitute in post-AGB stars}
\label{Sect:Li}

This is another exemplifying situation where inadequate atomic data
led astrophysicists to build fancy theories to explain a spurious
result, namely a high Li abundance in a very evolved class of stars
know as post-AGB stars.

On the AGB, stars with initial masses between $\sim$0.8 and 8 \Msun\ 
reach very large radii (several hundreds R$_\odot$) and low effective
temperatures ($4000 \ge T_{\rm eff} \ge 2000$~K;
Fig.~\ref{Fig:HR}). They moreover possess a deep convective
envelope. Post-AGB stars recently left the AGB and are about to become
white dwarfs (WDs), thus they are crossing the HR diagram from right
to left.

Lithium is a very fragile element. Its most abundant $^7$Li
isotope is destroyed by proton-capture reactions at temperatures of a
few 10$^6$~K \cite{Wallerstein-97}. During the first ascent of the
giant branch, the convective envelope deepens into layers devoid of $^7$Li,
so that the surface $^7$Li abundance of giant stars drops markedly. 
There is however the possibility to
produce  $^7$Li through the so-called Cameron-Fowler transport mechanism \cite{Cameron-Fowler-71}.
This mechanism is associated with ``hot-bottom burning'', a situation where the base of the
convective AGB envelope penetrates in layers  hot enough for $^3$He($\alpha,\gamma)^7$Be
to operate. The freshly produced $^7$Be is then quickly transported  to
cooler layers where its decay product, $^7$Li, can no longer be destroyed by proton captures
(\cite{Wallerstein-97} and references therein). This process is well
documented observationally by the detection of several Li-rich AGB
stars both in our Galaxy and in the Magellanic Clouds
(\cite{Reyniers02} and references therein). This hot-bottom process
is, however, only expected to be active in intermediate-mass stars
with initial masses around 4--5~\Msun.

\begin{figure}
\centerline{\psfig{file=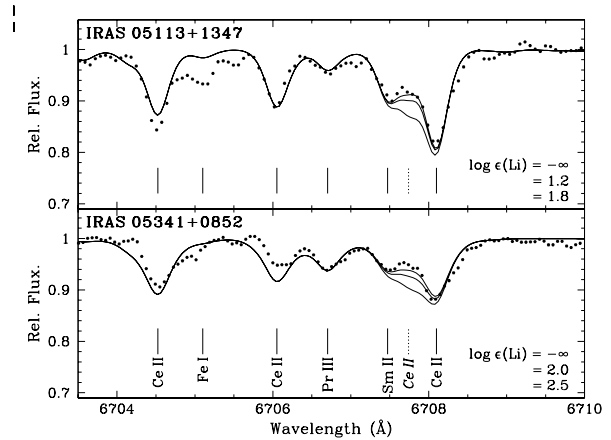,width=6cm,angle=270}}
\caption[]{\label{Fig:Reyniers}
Spectral synthesis of the region around the Li doublet in post-AGB
stars, 
revealing that 
the Ce~II line at 670.8099~nm provides a perfect match to the observed 
feature (dots) with no need for a high Li content  in those
stars. Note that yet another line seems to mutilate the Li line in
another class of stars enriched in s-process elements,  the 
so-called barium stars. That line was tentatively attributed to
Ce~II as well \cite{Lambert-93}. It is depicted by the dashed line and 
slanted font in the above figure. From \cite{Reyniers02}.
}
\end{figure}

The high Li abundances that were reported in {\it low-mass} post-AGB
stars came therefore as a surprise (\cite{Reyniers02} and references
therein).  Indeed, several such objects were described to be enhanced
in Li to such an extent that production of Li had to be invoked to
explain the derived abundances. A further anomaly was the fact that
the presence of Li in these stars was inferred from the famous Li
resonance doublet at 670.776 and 670.791~nm but,  in all stars, the
doublet seemed to be shifted redwards by about 0.02~nm compared to the
expected position based on other photospheric lines.  Motivated by
this unexplained shift in the Li doublet, Reyniers et
al. \cite{Reyniers02} investigated an alternative identification of
this line. They were greatly helped in this enterprise by the {\sc
DREAM} database ({\it Database on Rare Earths At Mons University}
\cite{Biemont-99,Biemont03}; see also Sect.~\ref{Sect:databases}). As stated in
\cite{Reyniers02}, \textsl{one of the problems dealing with strongly s-process enriched
objects is the lack of accurate atomic data of many s-process elements. Besides the more
general lack of accurate oscillator strengths, even the wavelengths of
the transitions of neutral and ionised s-process species are badly
known. The main purpose of the {\sc DREAM} database (...) is to
provide such an update concerning the radiative properties of
rare-earth atoms and ions.}  These authors soon realized that a Ce~II line at
670.8099~nm was an obvious candidate to replace the ``shifted Li
line.'' Spectral synthesis of the corresponding region leads to very
satisfactory fits to the observed spectra (Fig.~\ref{Fig:Reyniers}),
which led the authors to conclude \cite{Reyniers02} that  \textsl{ in none of
  the post-AGB objects there is evidence for in situ Li production and 
  there is no need to invoke special non-standard mixing 
  during the AGB evolution to explain the claimed high abundances of
  the fragile element in these stars. (...)
This result dramatically illustrates the importance of reliable and
complete line lists when doing analyses of these (...) objects (...).}

\subsection{Exploring the near-UV: The heavy elements}
\label{Sect:UV}

The {\it Goddard High-Resolution Spectrograph} \cite{GHRS} on board
the {\it Hubble Space Telescope} (until 1997), the {\it
  Ultraviolet-Visible Echelle Spectrograph} (UVES) on 
the {\it Very Large Telescope} \cite{UVES}, {\it
  HIRES} on Keck~I \cite{Vogt-94}, the {\it High Dispersion
  Spectrograph} on {\it Subaru} \cite{Noguchi-02} opened up the UV spectral 
window to high-resolution spectral studies.
Two classes of stars benefited from this new instrumentation:  
Cool, metal-poor stars enriched in heavy elements ($Z > 30$) and warm, main-sequence
chemically-peculiar (CP) stars, since they are
ideal candidates for exploring the wealth of heavy-element lines
present in the ultraviolet spectrum.

\subsubsection{CP stars}

\begin{figure}
\centerline{\psfig{file=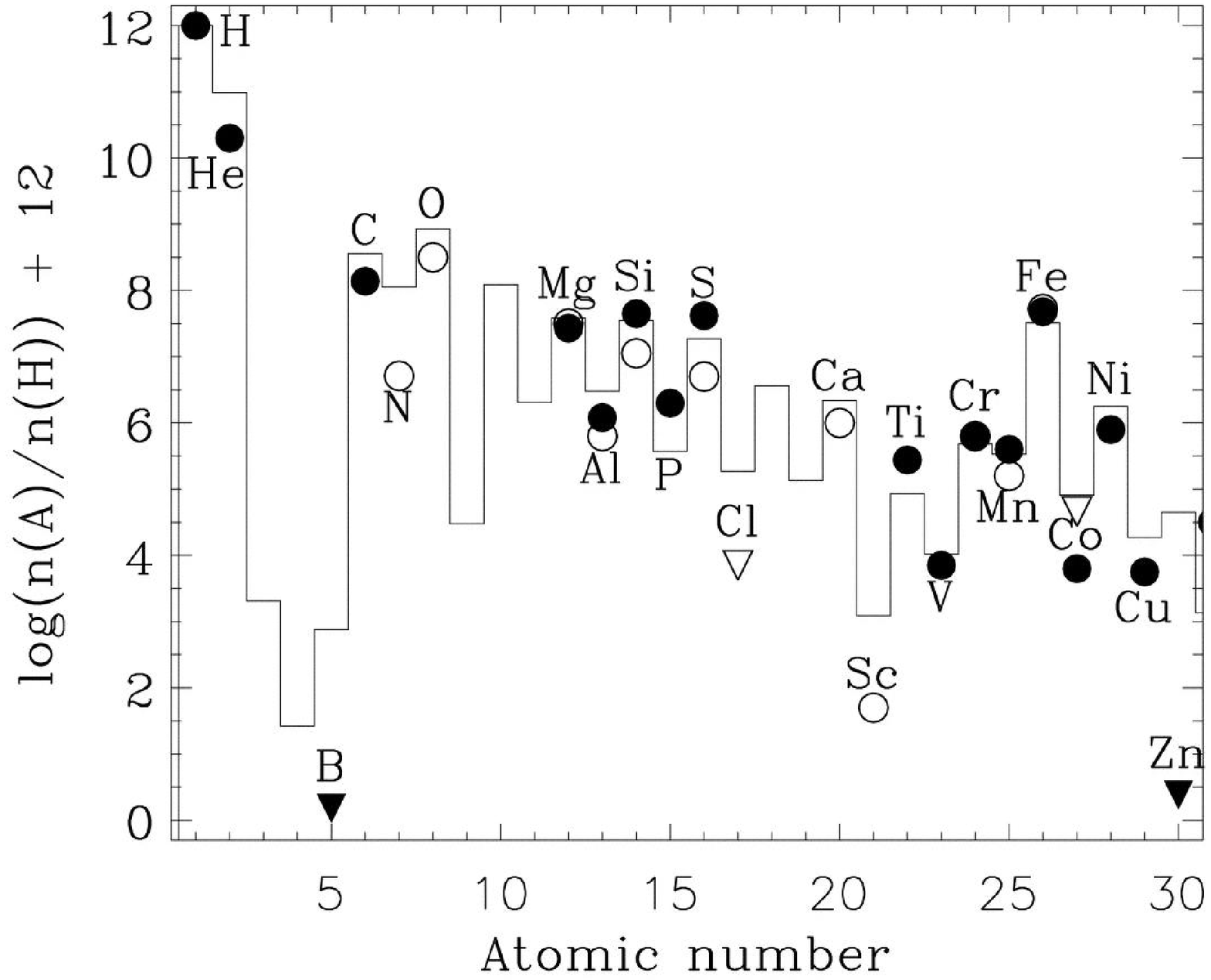,width=7.5cm}}
\centerline{\psfig{file=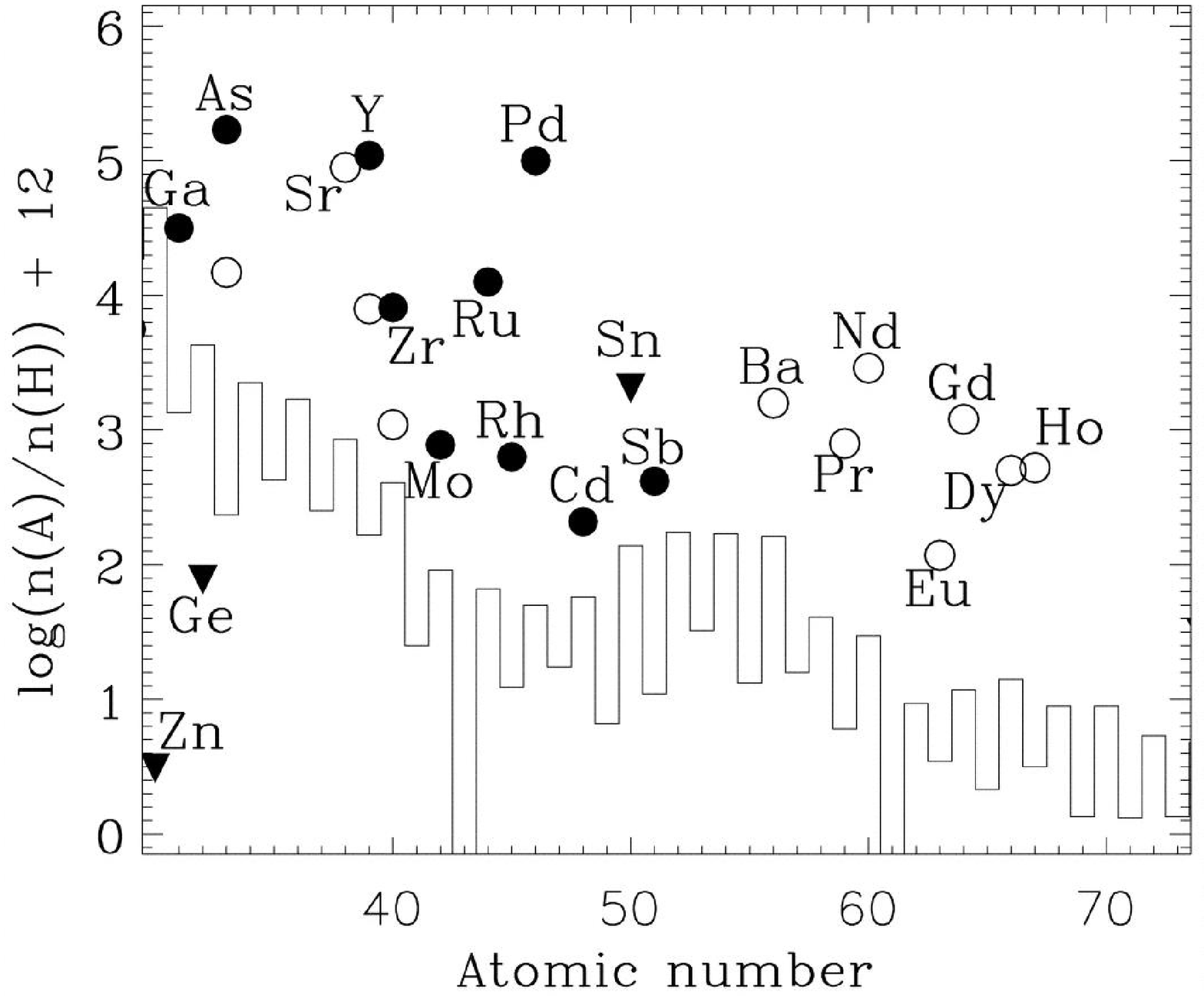,width=7.5cm}}
\centerline{\psfig{file=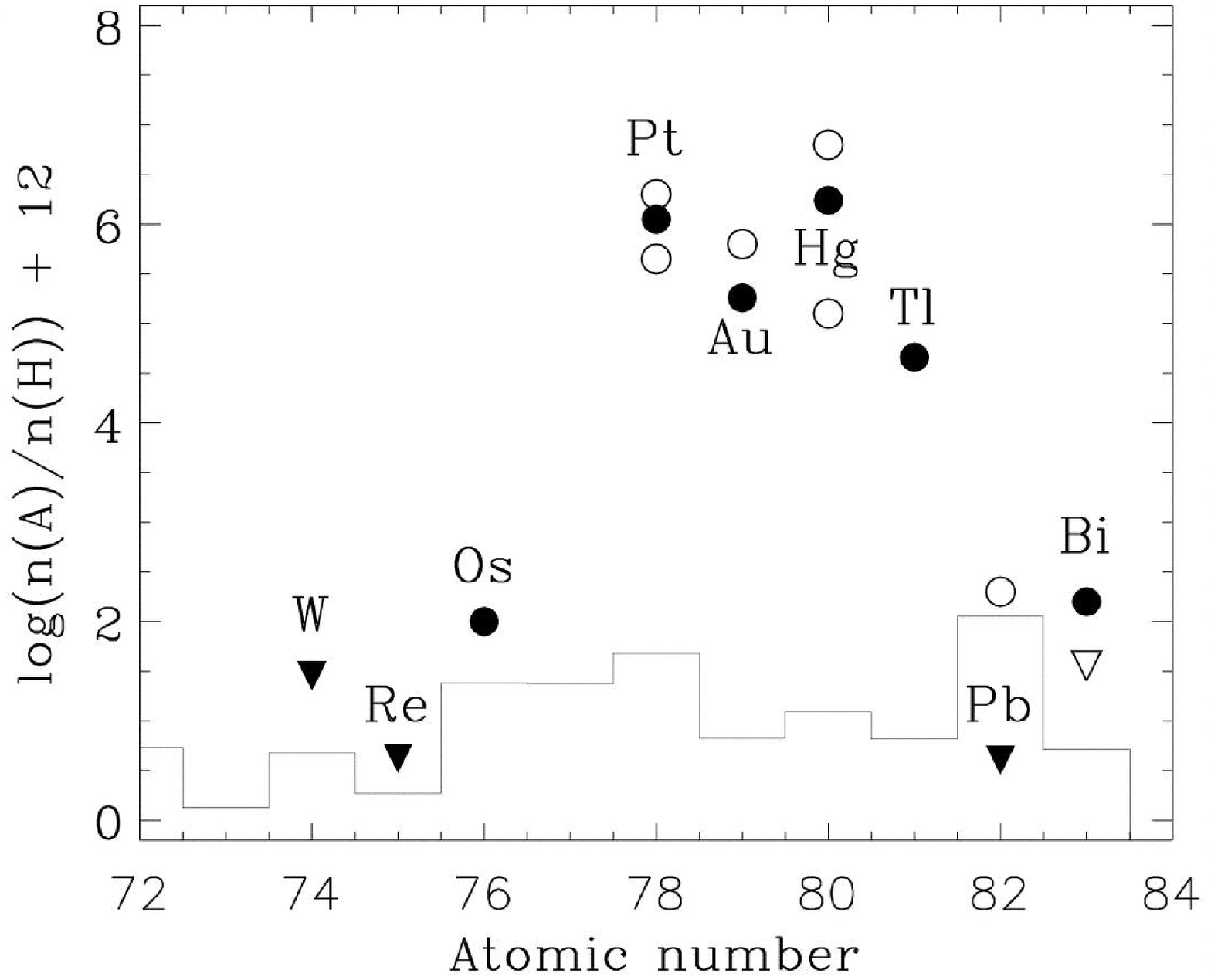,width=7.5cm}}
\caption{\label{Fig:chiLupi}
Abundances in the Bp star $\chi$~Lupi (filled symbols indicate that
the elemental abundance is derived from the main ionization stage; 
triangles indicate upper limits), compared with the solar system
distribution of elemental abundances (solid curve). Note especially
the large amplitude heavy-element peak around $Z = 80$ (platinum,
gold, mercury and thallium), and the depletion of boron ($Z = 5$) and
zinc ($Z = 30$).  This abundance pattern is not due to
nucleosynthesis, but rather to atomic processes (compare to
Figs.~ \ref{Fig:Cowan} and \ref{Fig:lead}). From \cite{Leckrone-99}. 
}
\end{figure}

Chemically-peculiar stars correspond to main-sequence stars of
spectral types B, A (denoted Bp, Ap, HgMn or Am stars). 
The chemical peculiarities that they display are not shaped by 
specific nucleosynthesis processes, but rather by physical processes
operating in the
hydrodynamically stable atmospheres
of these upper main-sequence, slowly-rotating stars. 
The atomic properties play a central
role in these processes, since the chemical fractionation occurring
in these atmospheres is governed by the competing effects of
radiatively-driven diffusion and gravitational settling \cite{Michaud-70,Smith-96a}.
Magnetic fields also play a role in some cases.
These processes lead to abundance anomalies, both depletions and enhancements, some of enormous
magnitude (up to factors of $10^5$) relative to normal stellar abundances, spanning essentially
the entire periodic table (Fig.~\ref{Fig:chiLupi}).

The challenge here is twofold: (i) realistic calculations of
radiatively driven diffusion require, besides sophisticated stellar
model atmospheres, a large database of atomic data covering many
ionization states of a particular element, and (ii) deriving
abundances from observed spectra to compare with model predictions
requires accurate atomic data for specific lines.  The problem
has been perfectly summarized in \cite{Leckrone-99}: \textsl{The
published database of energy levels, classified transitions,
wavelengths, oscillator strengths, and nuclear effects (hyperfine
structure and isotope shifts) in existence before the launch of the
GHRS in 1990 was badly deficient in its coverage and accuracy to allow
interpretation of the exquisitely detailed UV spectra of CP stars. The
existence of the GHRS observations of $\chi$~Lupi is now widely cited
by atomic physicists as a ``standard spectroscopic light source,''
because its UV spectrum displays transitions of so many rare elements
and also transitions of common elements (e.g., Fe~II), which are
difficult to observe in absorption in the laboratory. The large
database of atomic parameters [from Kurucz \cite{Kurucz95}] provides
the essential starting point for the synthesis of complex spectra. But
the active participation of atomic physicists in providing new, highly
accurate measurements or calculations for the plethora of ions
observed in the $\chi$~Lupi spectrum has been indispensable. The
synergy between atomic physics and astrophysics in this investigation
has led to new insights in both fields}
(\cite{Henderson-99,Johansson-98,Johansson-95,Leckrone-93,Leckrone-96,Palmeri-00,Wahlgren-01}
to quote only a few).

The detailed comparison between observed and predicted abundances of
the kind performed in \cite{Leckrone-99} allows one to check whether
radiatively driven diffusion and gravitational settling adequately
account for the entire range of phenomena displayed in CP stars. Some
systematics emerges \cite{Leckrone-99}: The occurrence of large
abundance anomalies is limited to elements whose initial abundance (in
the prestellar nebula) was low ($\log \epsilon < 5$), a simple result
of saturation in the absorption of radiative momentum. Moreover, clear
trends are observed for homologous elements (belonging to the same
column in the periodic table), in the expected way. But puzzles arise
as well: one is how---or whether---the above physical processes can
account for the isotopic anomalies observed in platinum, mercury and
thallium in $\chi$~Lupi for example \cite{Leckrone-99}.

\subsubsection{Cool, metal-poor stars}
\label{Sect:r}

\begin{figure}
\centerline{\psfig{file=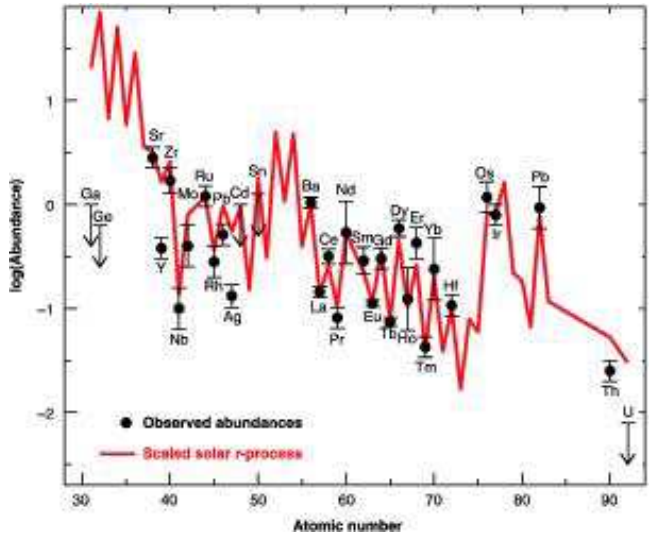,width=8.5cm}}
\caption{\label{Fig:Cowan}
Abundances in the very metal-poor star CS~22892-052 (black dots)
compared to the (scaled) solar-system r-process abundances. 
Compare to Fig.~\ref{Fig:lead}. From \cite{Sneden-03}.
}
\end{figure}

The near ultraviolet region (down to 360~nm) of cool, metal-poor stars has been
explored in \cite{Cowan96,Sneden96}, reporting the first detection in
low-metallicity stars of lines (see \cite{Sneden96} for a detailed list)
from ``exotic'' elements [like terbium (Tb, $Z = 65$), holmium
(Ho, $Z = 67$), thulium (Tm, $Z = 69$), hafnium (Hf, $Z = 72$), 
osmium (Os, $Z = 76$), platinum (Pt, $Z = 78$)] 
of special interest as tracers of r-process nucleosynthesis (The s-process will be addressed
in Sect.~\ref{Sect:lead}).  
The observations \cite{Cowan96,Hill-02,Sneden-98,Sneden96,Westin-00} 
of metal-poor stars with the various high-resolution
UV-visible spectrographs listed above led to the
identification of several 
low-metallicity
stars (CS~22892--052,  CS~31082--001, HD~115444, HD~126238) 
where the distribution of abundances for the
heavy elements ($56 \le Z \le 82$) is similar to the
solar-system r-process abundance pattern (Fig.~\ref{Fig:Cowan}).
These observations very clearly demonstrate (i) that the r-process abundance pattern seems to be ``universal'',
and (ii) that the r-process was already operating very early in the lifetime of the Galaxy, before these 
very old, metal-poor stars formed. The latter conclusion also supports
theory, which predicts that the short-living massive stars are the
site of operation of the r-process at the time of their Type~II
supernova explosion (see \cite{Wallerstein-97} and references therein). 

Even for metallicities as low as [Fe/H] $< -2.0$, the ultraviolet
spectrum of cool stars is packed with lines, and their correct
identification is quite challenging. 
Here again, the availability of extensive line lists like Kurucz's
\cite{Kurucz95} proved to be of
considerable interest, as stressed in \cite{Sneden-98,Sneden96}, but 
these authors warn that
\textsl{certainly there also  must  be (...) transitions [from heavy elements]
  lurking in the [stellar] 
  spectrum that simply have yet to be identified in the 
  laboratory} \cite{Sneden96}.


Experimental transition probabilities for  ``exotic'' elements like those listed above
were very sparse, and the
analyses of heavy-element-rich metal-poor stars  sparkled specific 
laboratory and theoretical 
studies performed by the Lund, Wisconsin and Mons-Li\`ege teams
(Sect.~\ref{Sect:loggf}).

\subsubsection{Lead stars}
\label{Sect:lead}

\begin{figure}
\centerline{\psfig{file=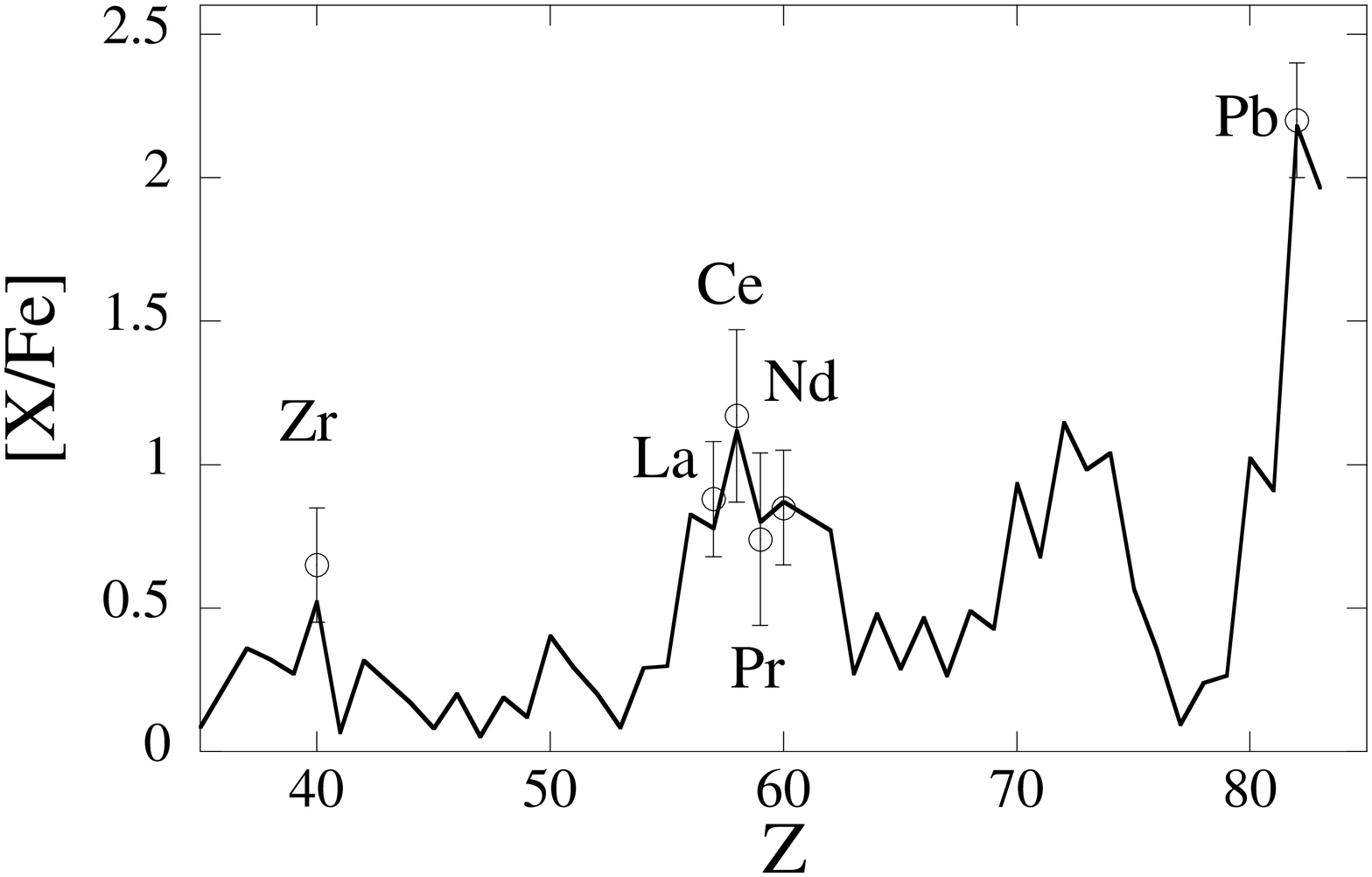,width=8.5cm}}
\caption{\label{Fig:lead}
Abundance pattern in the CH star HD~196944 (open circles)
compared to predictions \cite{Goriely-00} for the s-process operating in an
AGB star of the same low-metallicity ([Fe/H] $= -2.45$) as HD~196944
(solid line). Compare to Fig.~\ref{Fig:Cowan}. From \cite{VanEck-01}.
}
\end{figure}

Clear signatures from the operation of the s-process at low metallicities have been observed in
the so-called lead stars \cite{VanEck-01,VanEck-03}. These low-metallicity stars exhibit
overabundances of s-process elements along with large [Pb/heavy-s] ratios (where heavy-s denotes
any of the elements belonging to the second s-process peak).  
The very specific abundance pattern observed in lead stars (Fig.~\ref{Fig:lead}) agrees
with predictions for the operation of the s-process at low metallicities
\cite{Goriely-00}. Not all low-metallicity, s-process-rich stars are lead stars, however
\cite{VanEck-03}, and the origin of this variety, not predicted by the models, is still a
mystery. Solving this puzzle will undoubtedly represent an important step towards the
understanding of the operation of the s-process in stars.

\begin{figure*}
\begin{minipage}[b]{\columnwidth}
\caption{\label{Fig:CN}
\textbf{Right panel.} The U~II $\lambda 385.947$~nm line in the
metal-poor, r-process-rich star CS~22892-052, and the blending CN
line. From \cite{Gustafsson-01}.
\textbf{Bottom panels.}
The spectral region around the Pb~I 405.781~nm line (indicated by an arrow), 
observed at a
resolution of 135$\;$000 in the cool CH star V~Ari (thick solid line). 
The best-matching
spectral synthesis (thin solid line, 
adopting $T_{\rm eff} = 3580$~K, $\log g = 0.1$ ,
[Fe/H] $= -2.4$ and C/O $= 1.07$  \cite{VanEck-03})
is still poor, and the mismatch is likely due to remaining inaccuracies in the CN line list (dashed line). The
two rows of vertical ticks compare the intensities of the CN lines in our spectral
synthesis (lower row) and in a laboratory arc spectrum (upper row, from Davis, priv. comm.). It
is clearly seen that the wavelengths are not always correctly
predicted. The fact that some CN lines are missing may be due to the
different arc and stellar temperatures. The CN line list has been generated 
by B. Plez in a similar fashion as his TiO list \cite{Plez-98}, 
using the molecular data quoted in \cite{Hill-02}.   
[From S. Van Eck, priv. commun.].  
}
\end{minipage}
\begin{minipage}[b]{\columnwidth}
\centerline{\psfig{file=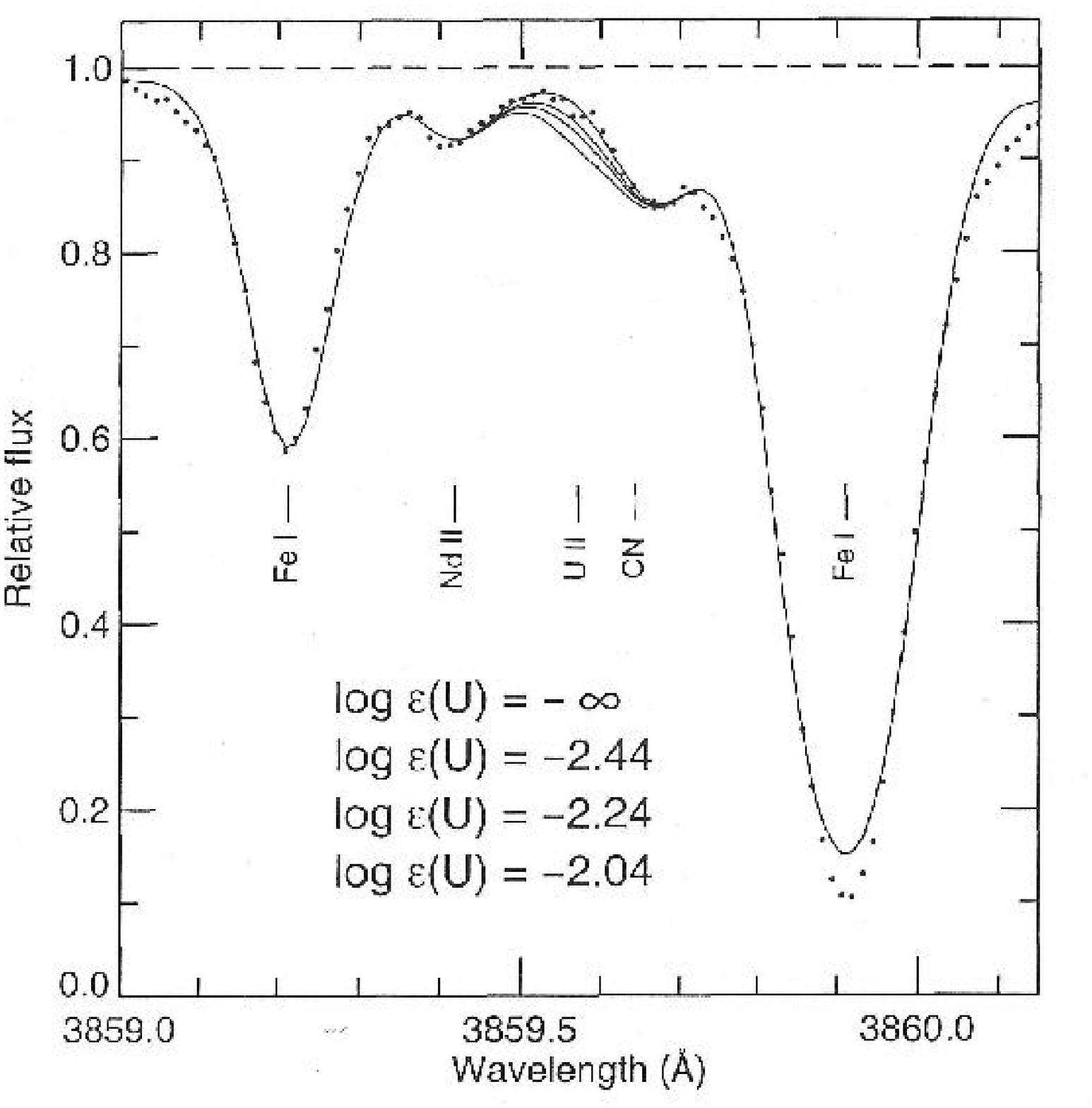,width=8.5cm}}
\end{minipage}
\centerline{\psfig{file=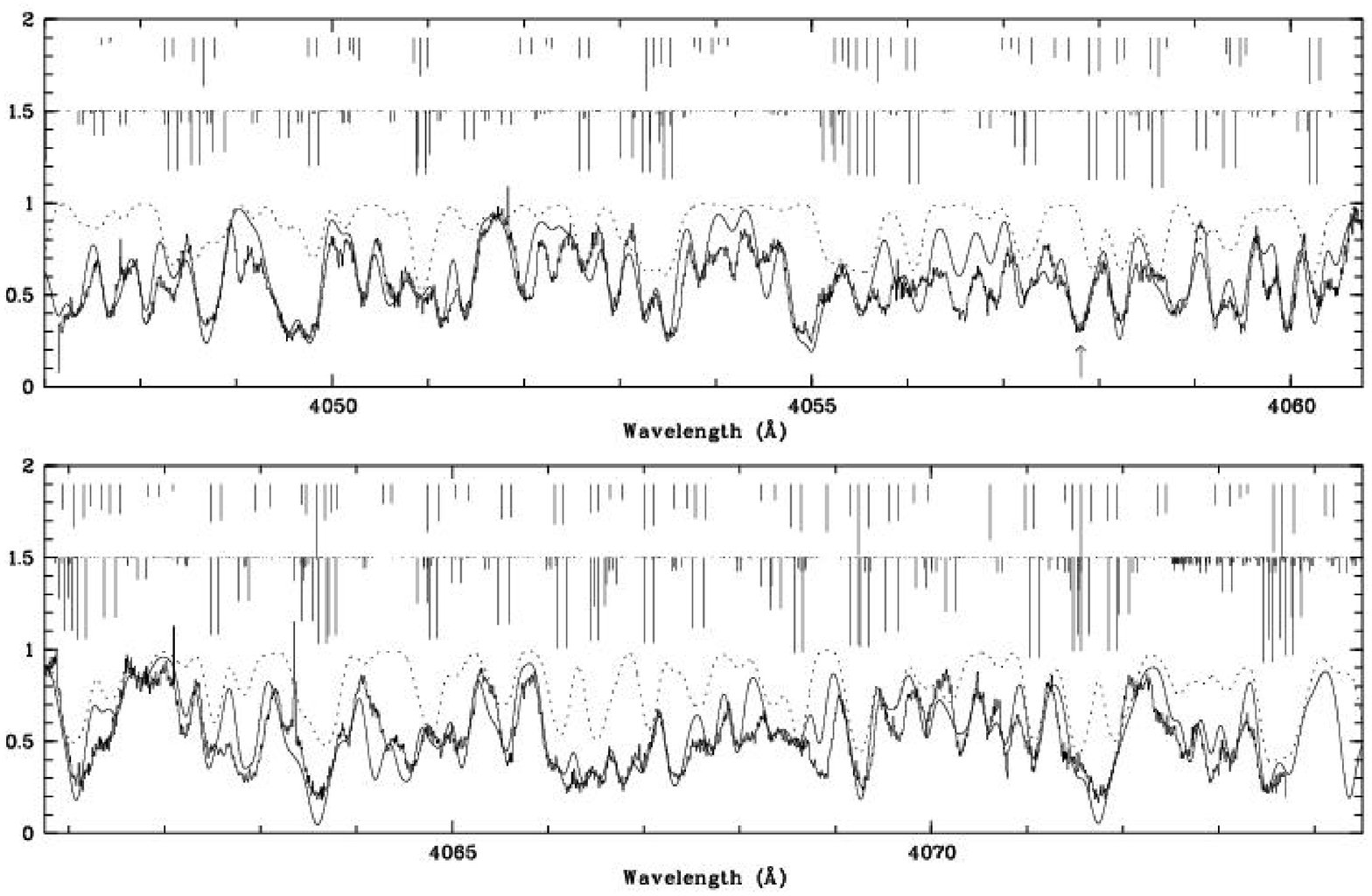,angle=270,width=17cm}}
\end{figure*}

The situation on the front of atomic and molecular physics is not free from problems either.
The Pb abundance has so far been derived from the Pb~I lines at
$\lambda$368.34 and $\lambda$405.781~nm 
\cite{Aoki2002, VanEck-03}. Accurate, laboratory oscillator strengths for these lines
have been derived by \cite{Biemont-2000}.
The major concern comes rather from the forest of CN lines in which the Pb~I $\lambda$405.781~nm
line is embedded when the star is carbon-rich. There is currently no entirely 
satisfactory line list available for the violet CN system. 
 The situation is made even more critical
by the fact that the CN lines in the region of the Pb $\lambda$405.781~nm line involve very
large
$J$ values ($60.5 \le J \le 80.5$ in the $R1$ and $R2$ branches of the  $v-v' = 0-1$ system), 
where models become generally less reliable. Fig.~\ref{Fig:CN}
reveals that the situation is far from being satisfactory when attempting spectral
syntheses in cool carbon stars.

The CN lines are also badly mutilating the spectral window where the
U and Th lines (U~II $\lambda385.957$~nm
and Th~II $\lambda401.9129$~nm) used for cosmochronology
(Sect.~\ref{Sect:oldest_stars}) are located. Gustafsson \cite{Gustafsson-01}
therefore writes: \textsl{Systematic efforts in molecular and
    atomic physics to trace and measure [the blending] features in the 
    U and Th line regions are of great [astrophysical] significance.}

\subsection{Dating the oldest galactic stars}
\label{Sect:oldest_stars}

The so-called r-process of nucleosynthesis
\cite{Goriely-99b} (see also Sect.~\ref{Sect:r}) is held responsible
for the production of the actinides, among which Th and U.
The first use of the radioactive decay of $^{232}$Th ($\tau_{1/2} =
14\times10^9$~y) as a chronometer has been
attempted by \cite{Butcher87,Butcher-88}. The stellar age is obtained by comparing the observed
Th/Eu (or Th/Nd) ratio with the predicted value at the stellar birth, the difference being due
to the radioactive decay of Th. 
This predicted value relies on both the  theoretical
production ratio at the nucleosynthesis (r-process) site and on the chemical evolution of the
Galaxy which controls the rate at which  the r-process nuclides are injected in the
interstellar medium (ISM).
Both ingredients are highly uncertain. 
The injection rate in the ISM may be a
strong function of time, depending upon the star formation rate and
the lifetime of the stellar r-process sites.
On the other hand, the stellar yields predicted by the
r-process are also quite uncertain, since the description of the r-process 
suffers from very many astrophysics and
nuclear physics problems \cite{Arnould01,Goriely-99b}. 
The physical conditions at the r-process site are badly known 
(as is the site itself!), as are the  nuclear
properties (e.g., masses, $\beta$-decay and electronic-capture rates) 
for the neutron-rich isotopes involved in the r-process and located far from the valley of
$\beta$-stability  \cite{Goriely-99a}. 
Moreover, the more widely separated the elements are in the periodic table, the more difficult
it is to provide reliable relative stellar yields from the r-process 
\cite{Goriely-99a,Goriely-99b}.
A chronometer much more robust than Th/Eu or Th/Nd  is therefore obtained by considering
the pair $^{232}$Th and $^{238}$U, since both nuclides are exclusively produced by the
r-process and their proximity in the periodic table makes  their
predicted yields much more reliable.

This chronometer has been
applied by \cite{Cayrel-01,Cayrel-01a} for the first time to the
r-process-rich, metal-poor star CS~31082-001, thanks to the detection of the U~II
$\lambda 385.96$~nm line. These authors (see also
\cite{Schramm-74} for a detailed description of the principles of nucleo-cos\-mo\-chro\-no\-lo\-gy)
have shown that the
ratio $\epsilon_*({\rm Th})/\epsilon_*({\rm U})$ in the star at the time
of the solar-system birth (which is easily derived from the
observed ratio and the age of the solar system, 4.6~Gyr) can be
expressed in terms of the same ratio 
$\epsilon_\odot({\rm Th})/\epsilon_\odot({\rm U})$ in the solar system at its birth:
\begin{equation}
\label{Eq:Th1}
\frac{\epsilon_*({\rm Th})}{\epsilon_*({\rm U})} = \frac{\epsilon_\odot({\rm
  Th})}{\epsilon_\odot({\rm U})} \;
\frac{\int_0^{t_{*-\odot}}h(t) \exp(t/\tau({\rm U})) \; {\rm d}t}
{\int_0^{t_{*-\odot}}h(t) \exp(t/\tau({\rm Th})) \; {\rm d}t},
\end{equation}
where $t_{*-\odot}$ is the time
elapsed between the birth of the star and that of the solar system, and $h(t)$ describes the common
injection rate of Th and U in the ISM by the r-process production sites. This function
encapsulates all our ignorance about these sites and their evolution in the Galaxy. 
Once $t_{*-\odot}$ has been derived, the production ratio at the r-process 
site, (Th/U)$_0$, may be derived from the relation:
\begin{equation}
\label{Eq:Th2}
\frac{\epsilon_*({\rm Th})}{\epsilon_*({\rm U})} = 
\left(\frac{\mathrm{Th}}{\mathrm{U}}\right)_0
\exp\left(-t_{*-\odot}\;\left(\tau^{-1}({\rm Th}) - 
\tau^{-1}({\rm U})\right) \right) 
\end{equation}
expressing that U and Th have decayed over the time separating the 
star birth from the solar system birth.

Equation~\ref{Eq:Th1} is especially interesting, since it makes it
possible to derive $t_{*-\odot}$
independently of any assumption of the r-process yields (Th/U)$_0$.
Depending on the assumption made for $h(t)$, Cayrel et
al. \cite{Cayrel-01,Cayrel-01a} find ages in the
range 10.5 to 12.6~Gyr for CS~31082-001, which appear quite reasonable.
Plugging these values in Eq.~\ref{Eq:Th2} then leads to values of the
r-process production ratio (Th/U)$_0$ which falls as well within the range of
reasonable theoretical yields.
 
This fair agreement should not occult, however, the fact that various
uncertainties plague the age determination. 
On top of the difficulties already discussed above and relating to the
galactic chemical evolution (through the uncertain function $h(t)$ entering
Eq.~\ref{Eq:Th1}) and to the r-process yields, there are problems on the
atomic physics side as well (see also \cite{Gustafsson-01}).   
In particular, two lines are blending the Th~II
$\lambda 401.9129$~nm line: a Co~I line at $\lambda 401.9126$~nm
\cite{Lawler90} and a 
V~I line at $\lambda 401.9136$~nm \cite{Pickering95}, 
not recognized in Butcher's pioneering work on the Th/Nd
chronometer \cite{Butcher87}.  
The Co~I line is more
sensitive to temperature than the Th~II line, so that its contribution 
to the blend will change with different stars.
With the Co~I line taken into account \cite{Lawler90}, ages
for the galactic halo on the order of 15--20 Gyr are obtained,
as compared to $<9.6$~Gyr as initially estimated by Butcher \cite{Butcher87}. The latter age is 
in blatant
contradiction with the large body of evidence for an older Galaxy. 
In yet another re-analysis by \cite{Morell92}, 
the hyperfine structure of the Co~I line was included as well, leading 
to the conclusion  
that \textsl{the large scatter prevents us from drawing any firm conclusions 
concerning the age of the Galaxy.} This is a long way from the
very small galactic age initially derived by \cite{Butcher87}, and
which attracted a lot of---undeserved---mediatic attention. Thus, once 
again, what once appeared to be an astrophysical puzzle vanishes when using    
the best available atomic and molecular data.

Yet, the re-analysis by Morell et al. \cite{Morell92} does not
constitute the end of this saga. 
These authors 
realized that there is yet another contributor to the Th~II-Co~I
blend. Based on a search in the large line list of Kurucz
\cite{Kurucz89}, they already suspected that the V~I $\lambda 401.9136$~nm 
line could be the culprit, 
although the $f$-value provided by Kurucz
precluded this line from having any impact on the Th-Co blend. 
The last word came from \cite{Pickering95}
who measured  $\log gf = -1.30$, about five times larger than the
value of $-2.0$ from Kurucz. 
This example 
illustrates the importance of large
line lists like Kurucz's for performing exhaustive line searches, 
but reveals as well the danger of using blindly data from bulk computations 
for specific lines (more on this caveat in Sect.~\ref{Sect:databases})! 

In metal-poor stars enriched in carbon and neutron-capture elements, 
further contamination of the Th~II line by the Ce~II $\lambda
401.9057$~nm line occurs \cite{Johnson01,Sneden96}, as well as by
several $^{13}$CH lines from the $B\Sigma^- - X^2\Pi(0,0)$ band
\cite{Johnson01,Norris97b}. 


These extensive data on the blending lines, plus the recent
measurement of the oscillator strengths of the Th~II $\lambda
401.9129$~nm and U~II $\lambda 385.96$~nm lines
\cite{Biemont02,Nilsson02a,Nilsson02b}, 
should reduce the
uncertainty sources from atomic physics to a minimum. These atomic data 
are now routinely included in the latest age determinations based on
the Th chronometer (e.g., \cite{Johnson01}).


\subsection{Exploring the near-IR: the {\it ISO} legacy}
\label{Sect:ISO}

The {\it Infrared Space Observatory} (ISO) was equipped with two
spectrometers, the {\it Short-Wavelength Spectrometer} (SWS, covering
the range 2.38--45.2~$\mu$m) and
the {\it Long-Wavelength Spectrometer} (LWS, covering
the range 43--200~$\mu$m) \cite{ISO}.  
The ISO spectral data triggered studies of relevance to
the atomic physics community. For example, the accurate flux
calibration of SWS requires the observation of spectra of stellar
templates, and their comparison with flux-calibrated synthetic
spectra \cite{Decin-00}. 
However, our ability to construct reliable synthetic spectra
in this IR wavelength domain is still incomplete, and must be improved 
with the ISO data themselves. To achieve convergence of this iterative
process, it is essential to use reliable atomic data. Synthetic spectra
computed with state-of-the-art {\sc MARCS} atmospheric models 
\cite{Gustafsson-75} for warm stars (A0--G2) reveal that the existing
atomic data (Kurucz \cite{Kurucz95}, Hirata \& Horaguchi
\cite{Hirata-95}, Opacity Project \cite{Seaton-94}, VALD
\cite{Kupka-99}, van Hoof \cite{VanHoof-98}; see
Sect.~\ref{Sect:databases}) 
are unsatisfactory in the wavelength range of SWS
\cite{Decin-03}. Sauval (priv. comm.) has therefore constructed a new
atomic infrared line list with astrophysically-calibrated oscillator
strengths (more on this in Sect.~\ref{Sect:Sauval}).
The improvement achieved with these new oscillator strengths when fitting  
synthetic spectra to ISO-SWS spectra is
spectacular, as revealed by Figs.~5 and 6 of \cite{Decin-03}.


\subsection{Exploring the far-IR: the {\it Herschel} promise}
\label{Sect:Herschel}

The exploration by ISO of the 2--200~$\mu$m range
 will be pursued and extended by the {\it Herschel Space
 Observatory} (to be launched in 2007) 
exploring the 60--670~$\mu$m range with
high spectral resolution capabilities \cite{Herschel}.
Observations with the {\it Herschel Space
 Observatory} can be expected to lead to the detection of thousands of 
new spectral features due to a variety of simple and complex molecules.
Identification, analysis and interpretation of these features in terms 
of the physical and chemical characteristics of the astronomical
sources will require laboratory measurements and theoretical line strength and
collisional excitation studies on species of astrophysical relevance.
Of particular relevance are: (i) fine structure transitions of atoms
and atomic ions \cite{Castro-01}
(ii) pure rotational transitions in the ground state
as well as in the excited vibrational states of small molecules and
radicals, including in particular $\mathrm{H_2O, H_3O^+, CH^+, OH, NH, 
  CH, SH, LiH, HCN, CH_2}$,\\
$\mathrm{HCO^+, NH_2, C_2H, HF,}$ etc. (iii)
rotational-torsional spectra of small internal rotors like
$\mathrm{H_2CO, CH_3OH, CH_3CN,}$\\ $\mathrm{CH_3SH, CH_3COOH,}$ etc.
  
More details about the work that needs to be done in this context may
be found in the Herschel ``white book''
at {\tt
  http://www.astro.rug.nl/$\sim$european/}.
For example, existing lists of atomic lines (like\\ {\tt 
http://www.mpe-garching.mpg.de/iso/linelists/}\\
{\tt FSlines.html}) should be
extended to cover the full {\it Herschel} spectral range.

\subsection{The pulsation mechanism of $\beta$ Cephei stars}

The problem of the pulsation mechanism of $\beta$ Cephei stars
remained unsolved for more than 30 years, and its solution
\cite{Cox-90,Cox-92,Moskalik-92} 
came from atomic physics. It thus provides yet another illustration of the
cross-fertilization between this discipline and astrophysics.

The variable stars of $\beta$ Cephei type are a group of early B stars
which exhibit short-period variations of brightness, radial velocity
and line profile \cite{Sterken-93}. These variations are clearly due
to pulsations, with periods ranging from about two to 
seven hours. But the excitation mechanism of these pulsations remained 
elusive. This excitation mechanism
must feed as much pulsational energy to the star from its total energy 
supply as is lost through damping during pulsation. A very effective
excitation mechanism is the so-called $\kappa$-mechanism
\cite{Baker-62}, which relies on the fact that certain layers of the
stars may become opaque at the time of maximum compression. If this
happens, the radiative energy becomes ``dammed up'' inside, causing the star to
expand again after  compression, and thus acting to drive the
pulsations. To be operative, this mechanism requires the opacity to
increase sharply with compression. This situation arises in regions of 
partial ionization, and the partial ionization zone He$^+$ -- He$^{++}$
has been shown to be responsible for the $\kappa$-mechanism operating
in the classical Cepheids and RR Lyrae variables \cite{Baker-62}. That 
mechanism cannot, however, be effective in the much hotter
$\beta$~Cepheid stars \cite{Christy-66}. 
The OPAL opacities \cite{Cox-92,Rogers-Iglesias-92,RogersIglesias98} 
provided the
solution to this puzzle by revealing an ``opacity bump'' due to the
sudden appearance of a tremendous number of iron lines as the 
temperature rises above $\sim10^5$~K. This opacity bump is
responsible for the operation of the $\kappa$-mechanism in
$\beta$~Cepheid variables \cite{Cox-90,Cox-92,Moskalik-92}. It should be
stressed that this sharp increase in opacity was found only after 
the spin-orbit interactions were properly included 
in the OPAL calculations  \cite{Iglesias-92}.    

\subsection{Modeling the atmospheres of brown dwarfs and very
  low-mass main-sequence stars}

The crop of extremely cool stars and substellar objects has been
meager until very recently, when marked improvements in detection
ability have finally started to yield a rich harvest
\cite{Allard97,Nakajima95,Oppenheimer95,Rebolo95,Ruiz97}. 
These very low-mass and very cool 
main sequence stars extend from the usual class of M dwarfs ($4000 \ge T_{\rm eff} \ge
2000$~K)   to brown dwarfs of newly-defined spectral classes~L and T. Class~L 
($2000 \ge T_{\rm  eff} \ge 1500$~K) is defined by the absence of TiO and
VO bands in the visible spectrum, which is dominated instead by bands of metal
hydrides like FeH, CrH and MgH. 
Class~T ($1500 > T_{\rm
  eff}$(K)) corresponds to the appearance of CH$_4$ bands in the
spectrum. The L and T classes are the first additions to the standard 
spectral-type 
sequence in over 60 years, and correspond to the cooling sequence for
a typical brown dwarf.  

M dwarfs span a mass range from 0.5~\Msun\ (M0, or $T_{\rm eff} \sim  3800$~K) down to the limit of
0.075~\Msun\ (M10, or $T_{\rm eff} \sim  2000$~K) 
corresponding to the lowest stellar mass which can
sustain hydrogen burning. Below this limit and down to 0.013~\Msun, 
deuterium-burning is operating, and defines the brown-dwarf regime.

Below $T_{\rm eff} \sim 1800$~K, complex O-rich compounds condense in
the atmosphere (corundum $\mathrm {Al_2O_3}$, enstatite $\mathrm
{MgSiO_3}$, forsterite $\mathrm {Mg_2SiO_4}$, perovskite $\mathrm
{CaTiO_3}$ etc.) and may form dust
clouds at different heights in the atmosphere depending on their
precise condensation temperature.

Advances in the atmospheric modeling of very cool stars have been slowed
by the twin bottlenecks of (i) incomplete molecular opacity databases 
(especially for $\mathrm {H_2O}$, $\mathrm{CH_4}$ and CaOH 
\cite{Kurucz02},  the latter having several strong bands
in the visible spectrum) and (ii) the inability to handle convection 
rigorously (but see
Sect.~\ref{Sect:3D} for a description of recent progresses in this field). Once these problems are addressed reasonably well, we
still face other challenges: incorporating the effects of photospheric 
grain formation, chromospheres, magnetic fields, departures from local 
thermodynamic equilibrium, spatial variations in atmospheric structure 
due to starspots, and cloud formation.
Grains play a very important role in these atmospheres, 
and possibly lead to weather patterns (grains are raining!). 
Dusty condensate clouds play an
important role in controlling the thermal emission
spectra of the warmest brown dwarfs, and much work is still needed to
improve our knowledge of the optical properties of the involved dust grains.


\section{Atomic physics and spectroscopy: A success story}
\label{Sect:success}

\subsection{Solar oscillator strengths for infrared lines}
\label{Sect:Sauval}

Kurucz \cite{Kurucz02a} has stressed that only about half the lines of
the solar spectrum are correctly identified. The situation was even worse
in the infrared. Geller \cite{Geller92} made a first attempt to improve
the situation using the ATMOS high-resolution solar spectrum
\cite{Farmer89}.  Sauval \cite{Decin-03} made the final step forward by
critically reevaluating Geller's identifications, and by providing
oscillator strengths (for atomic lines  in the 1--40~$\mu$m
solar spectrum) 
calibrated on the Sun using the Holweger-M\"uller model. 

\subsection{Laboratory measurements of transition probabilities}
\label{Sect:loggf}

The availability of tunable lasers and Fourier Transform
Spectrometers, coupled to the astrophysical impetus imparted by 
some of the hot
topics mentioned in Sect.~\ref{Sect:hot}, 
accounts for the wealth of accurate ($\pm5\%$) oscillator strengths
that have become available recently. They were derived by combining 
measurements of radiative
lifetimes (from laser-induced fluorescence) and of branching fractions
(from relative line intensities).

The {\it University of Wisconsin} team provided new transition probabilities 
and hyperfine structure data for 
(i) La~II \cite{Lawler01a}, leading to a revision of the solar La
abundance; (ii) Dy \cite{Wickliffe00}; (iii) Tb~II \cite{Lawler01b};
(iv) Eu~I \cite{DenHartog02} and Eu~II \cite{Lawler01c}, 
leading to a revision of the solar Eu abundance. 

The group at {\it Lund University} has provided experimental oscillator
strengths for the {U}~{II} and {Th}~{II} lines
\cite{Nilsson02a,Nilsson02b} of interest in
dating the oldest stars in the Galaxy (see
Sect.~\ref{Sect:oldest_stars}), as well as for ions visible in the
ultraviolet spectra of CP stars  (Sect.~\ref{Sect:UV}), like 
Ta~II, W~II, Re~II \cite{Henderson-99},  and Bi~I, II and III 
\cite{Wahlgren-01} to cite just a few.

The Lund team is also involved in the {\sc FERRUM} project 
\cite{Johansson02a,Johansson02b}. This project aims
at measuring oscillator strengths for
transitions from Fe~I and II in the ultraviolet and optical wavelength
regions involving 
energy levels with an extended span in excitation energy. This project
should remedy to the very unsatisfactory situation caused by
the very small number of Fe (especially Fe~II) lines 
with accurately measured $f$ values, as stressed in
Sect.~\ref{Sect:Fe} (see also \cite{Fuhr-88,Lambert-96}). This led 
Grevesse \& Sauval \cite{Grevesse99} to 
state: \textsl{In each \AA\ of the solar spectrum, we find an iron
  line. At the end of this Millenium, it is very disappointing to
  realize that accurate transition probabilities are known for only a
  minority of these lines.} This plea is being answered by the Lund
team and their {\sc FERRUM} project 
\cite{Karlsson-01,Pickering-02,Pickering-01}, so that the situation
regarding Fe~II is improving quite substantially (see also
Fig.~\ref{Fig:FERRUMKurucz}). 
 
The group at the {\it Universities of Mons} and {\it Li\`ege} 
focuses on rare earths (see
Sect.~\ref{Sect:databases} and \cite{Biemont-99}; also \cite{Biemont03} 
for a review of the lanthanides $57 
\le Z \le 71$, with an impressive list of references providing 
transition probabilities, and isotopic and
hyperfine splittings), but other elements (e.g., Pb
\cite{Biemont-2000}; Sect.~\ref{Sect:lead}) have been studied
occasionally as well. 

\subsection{Completeness of partition functions}

The degree to which the level structure of a given ion is complete is
critical to the reliability of the partition function. Inaccuracies in
the partition function systematically affect abundances derived from all
spectral lines of the considered ion. Much progress has been made in
recent years towards the completeness of the level structure for many
ions of astrophysical interest, as reviewed by \cite{Cowley03}, and
{\sl abundance workers would do well to make sure that they are using
the most complete and accurate material} \cite{Cowley03}. 

\subsection{A theory for Van der Waals broadening by neutral hydrogen atoms}
\label{Sect:OMara}

The Van der Waals broadening due to atomic collisions 
with neutral hydrogen is the dominant
source of line broadening in cool-star atmospheres. 
For decades, the only available model of collisional damping of atomic
lines was encapsulated in the Uns\"old formula \cite{Unsold-55}, strictly valid
for hydrogenic atoms collisioning with neutral hydrogen. 
For non-hydrogenic atoms, this treatment yields damping 
constants that are too small (by factors 2--10), so that 
a ``fudge-factor'' or ``damping enhancement factor''  
was applied to the Uns\"old
formula to estimate  the damping constant. Decisive progress was achieved
recently in this field 
\cite{Anstee95,Barklem98b,Barklem97,Barklem98a}, since a good model for
computing accurate cross sections for the broadening of s--p, p--s, p--d,
d--p, d--f and f--d transitions of neutral atoms by collisions with
neutral hydrogen is now available. This allows for the first time to get
rid of one of the most uncertain parameters of stellar spectroscopy, the
damping enhancement factor (The next section describes how another long-standing
{\it ad hoc} spectroscopic parameter, the microturbulent velocity, is about to
disappear as well!).

\begin{figure}
\centerline{\psfig{file=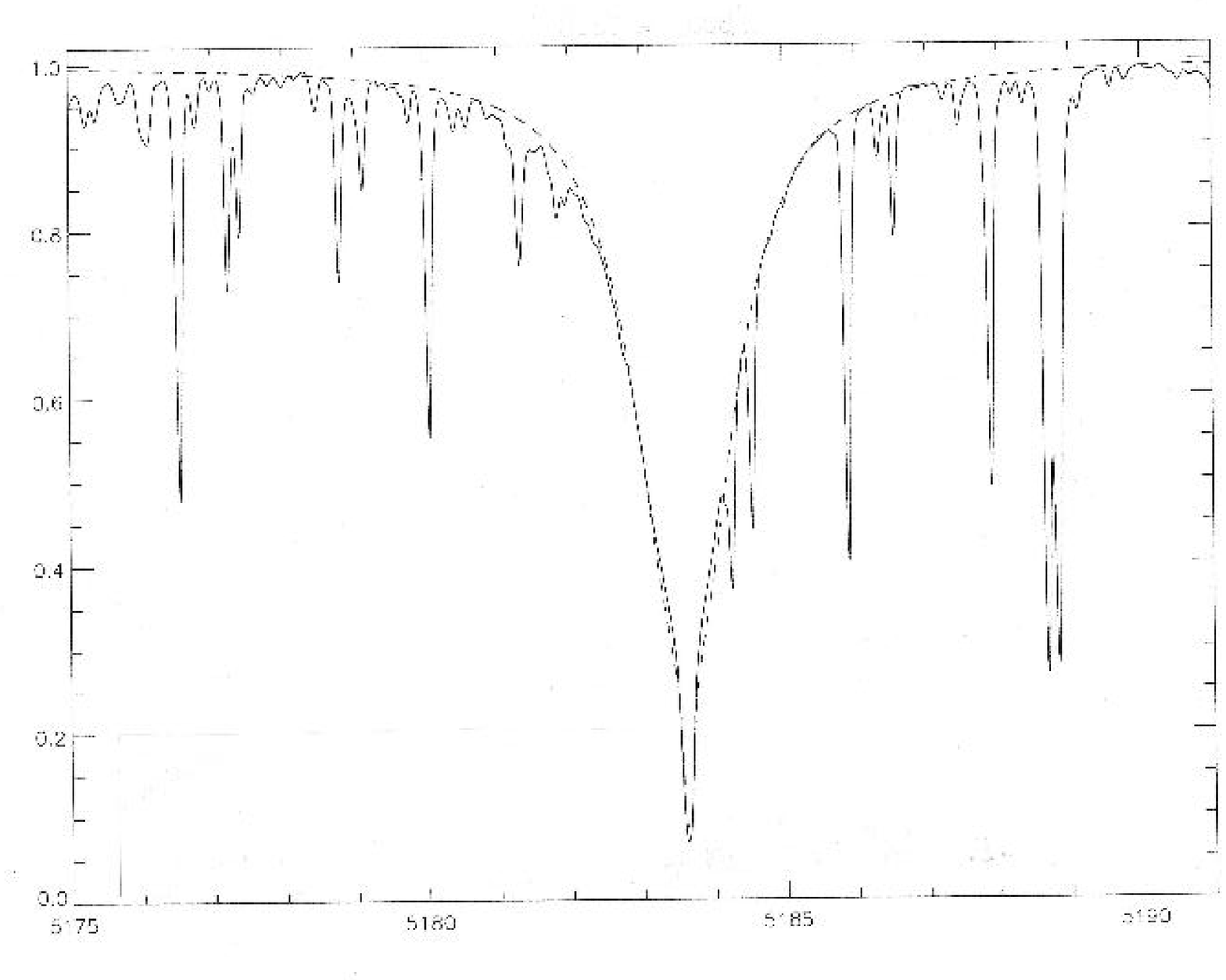,width=7cm,angle=90}}
\centerline{\psfig{file=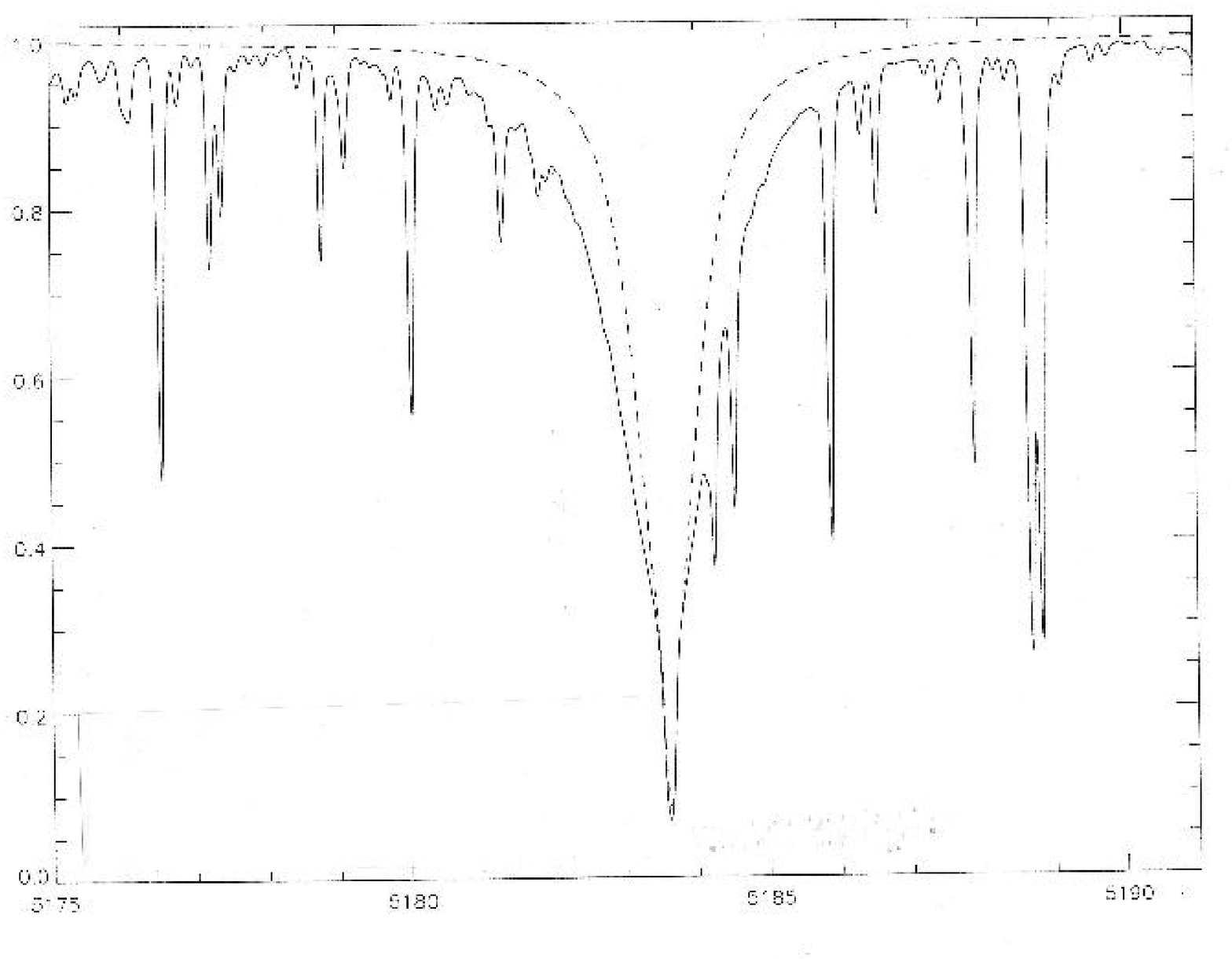,width=7cm,angle=90}}
\caption{\label{Fig:Grevesse}
Spectral syntheses of the Mg~I~$\lambda$518.3~nm line in the Sun, with the
Uns\"old model with no damping enhancement factor  (lower panel), and with the new damping
model  (upper panel).  The  Holweger-M\"uller
model revised according to 
\cite{Grevesse99}  has been used. The improvement is indeed 
spectacular. An abundance of
$\log
\epsilon(\mathrm{Mg}) = 7.60$ provides the best fit in the right wing
(around $\lambda518.5$~nm), as compared to the meteoritic value of  
$\log \epsilon(\mathrm{Mg}) = 7.57\pm0.01$.  From Grevesse (priv.
comm.). }
\end{figure}

The availability of this theory of collisional broadening makes it
thus possible to use the wings of very strong lines as abundance
indicators, a very unusual approach thus far, but which offers the big 
advantage that accurate oscillator strengths are generally available
for such lines. An illustration of this approach is presented in
Fig.~\ref{Fig:Grevesse}, which displays spectral syntheses of the Mg~I
$\lambda$518.3~nm line in the Sun, with the Uns\"old model and
with the new one. The improvement is
indeed  spectacular.

\subsection{3-D modeling of atmospheric convection: microturbulence
  is gone!}
\label{Sect:3D}

\begin{figure}
\centerline{\psfig{file=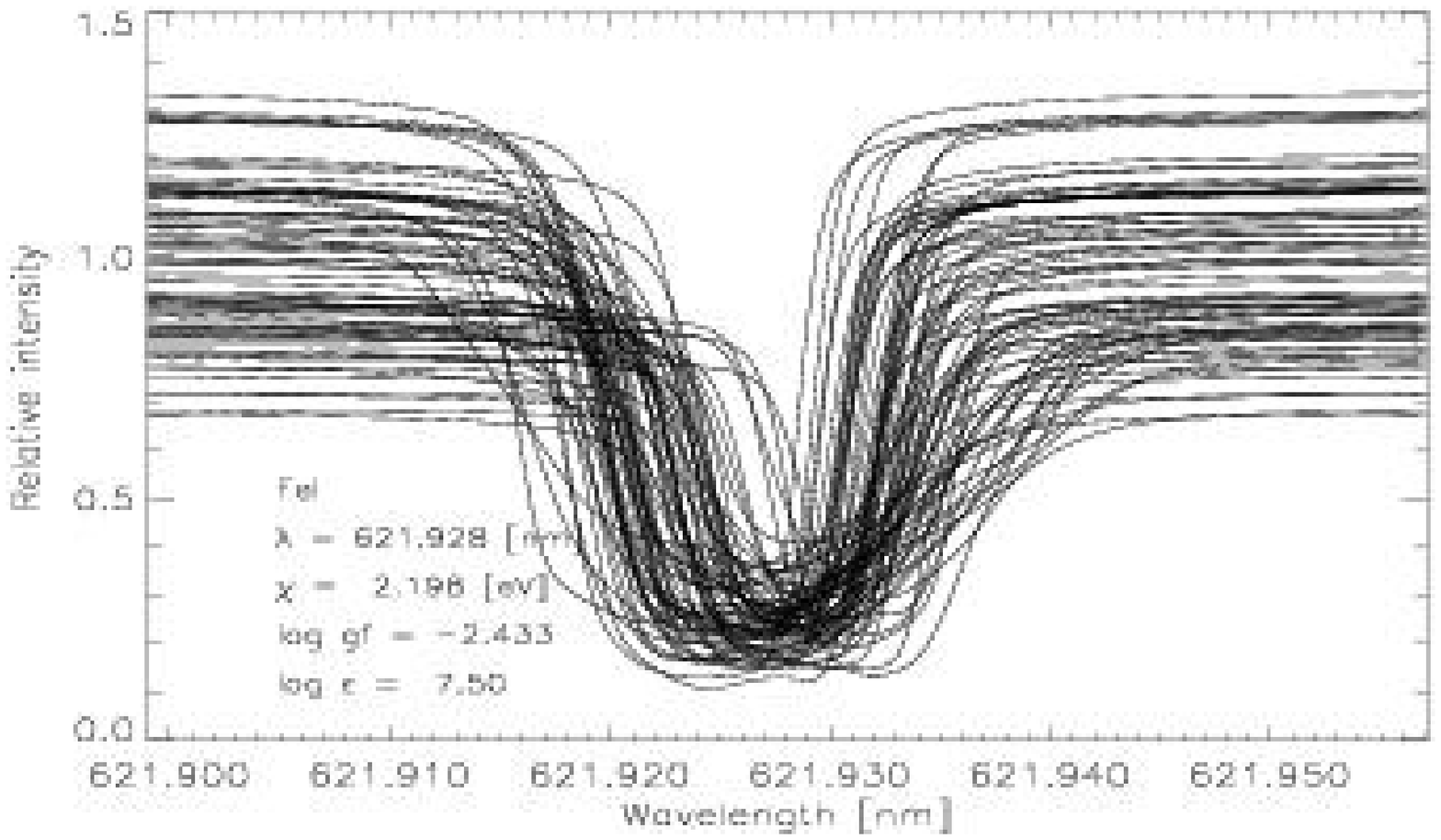,width=8.5cm}}
\centerline{\psfig{file=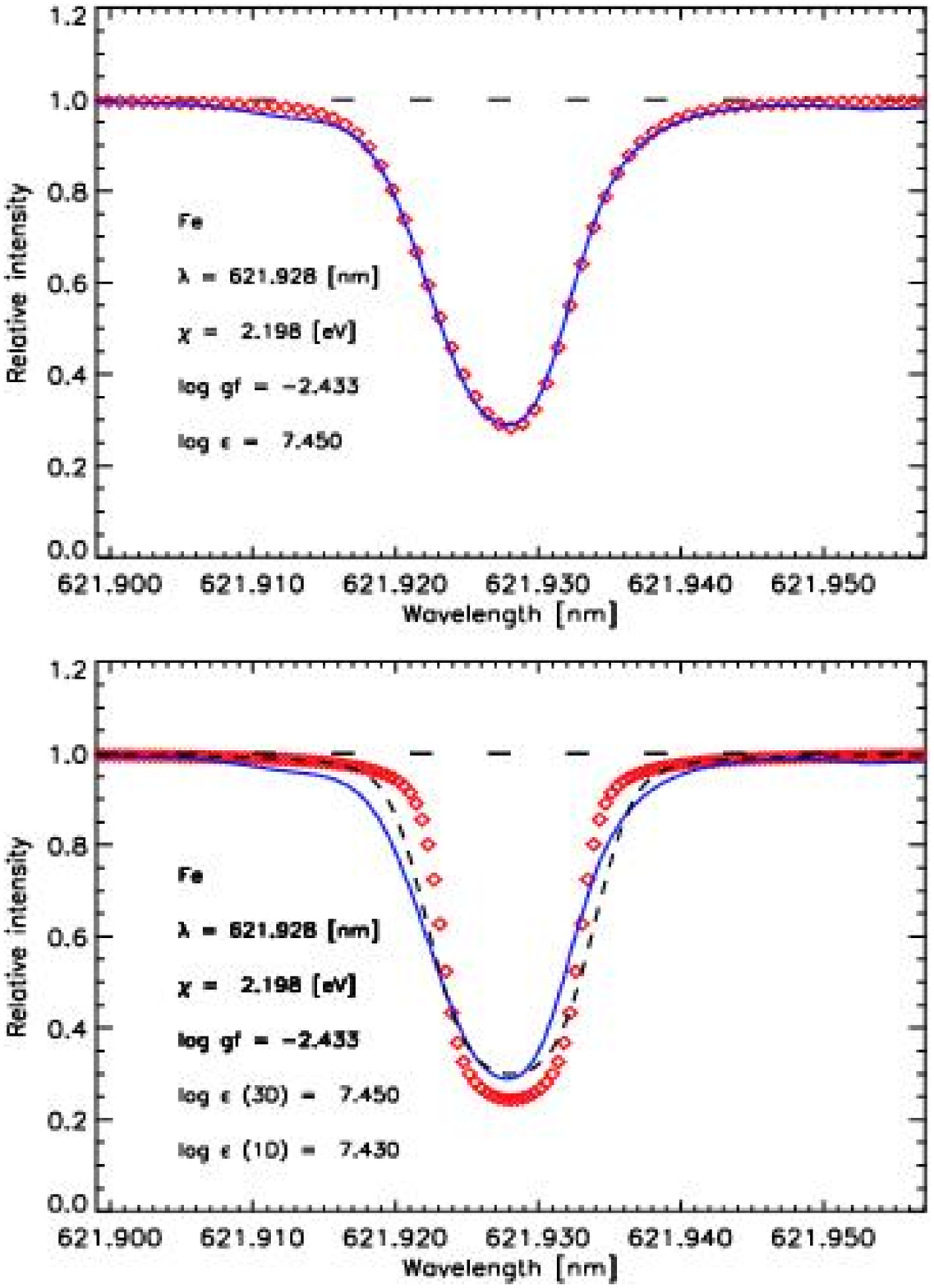,width=8.5cm}}
\caption{\label{Fig:Fe-3D}
{\it Upper panel:} 
A selection of line profiles for the Fe~I 621.928~nm at different locations  over the solar
disk at a specific instant, obtained from 3-D self-consistent simulations of the solar
convective atmosphere. Regions with upward-moving matter typically have stronger,
blue-shifted lines with higher continuum intensities than the downflows. 
{\it Middle panel:} Spatial (disk-limb) and temporal average of the line shapes presented
in the upper panel (diamonds), and comparison with the observed profile in the Sun (solid line).
{\it Lower panel:} Same as middle panel, when artificially removing all Doppler shifts
arising from the convective motions (diamonds). Also shown (dashed line) is an optimized
1-D line profile with micro- and macroturbulence yielding the same equivalent width as
the 3-D profile shown in the middle panel. 
 From
\cite{Asplund-03}.}
\end{figure}

The atmospheres of stars of solar and later types are convective (as revealed in the case of
the Sun by its surface granulation). The accompanying matter motions broaden the spectral
lines. In one-dimensional (1-D) analyses of stellar atmospheres, two {\it ad hoc}  parameters
are introduced to account  for this line  broadening, namely the micro- and macroturbulence  
\cite{Worrall-73}. They are intended to describe the line broadening associated with
mass motions on length scales less than and larger than a unit optical depth, respectively. 
Such a simplistic division is clearly artificial since the motions occur on a range of
scales. Almost all abundances for late-type stars derived so far have been obtained in this
framework, which has clear shortcomings, like its inability to reproduce the
observed asymmetric line shapes (solid line in the lower panel of
Fig.~\ref{Fig:Fe-3D}, to be compared with the diamonds representing the line profile obtained
with 1-D modeling).

\begin{figure}
\centerline{\psfig{file=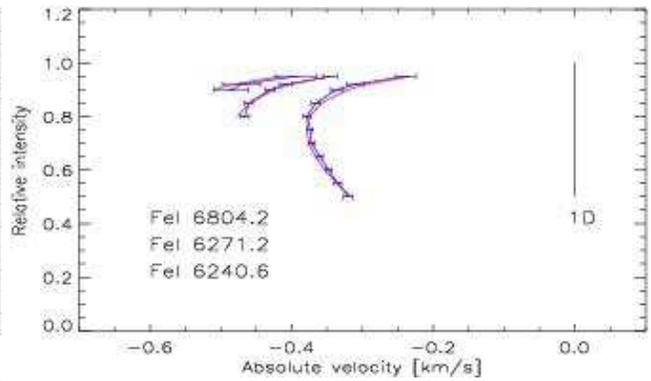,width=8.5cm}}
\caption{\label{Fig:bisector}
Examples of predicted (solid lines) and observed (horizontal error bars)
spectral-line bisectors of three Fe~I lines in the Sun. The agreement even on an {\it absolute}
wavelength scale is clearly highly satisfactory, as compared to  1-D profiles which are
purely symmetric (vertical line to the right).
 From \cite{Asplund-03}.}
\end{figure}

With the advent of supercomputers, self-consistent, three-dimensional,
radiative-hydrodynamical simulations of the convective atmosphere of solar- and late-type
stars have become feasible (\cite{Nordlund-90}, and, more recently \cite{Asplund-00}).
These simulations achieve a spectacular agreement between the observed and computed line
shapes in the Sun, especially in terms of  asymmetries  and absolute velocity shifts
(middle panel of Fig.~\ref{Fig:Fe-3D}, and Fig.~\ref{Fig:bisector}). 
Such a detailed agreement
ensures that these simulations reproduce the details of the velocity field, both in terms of depth
and surface structure (i.e., granulation). These simulations also enlighten the inadequacy of the
micro-  and macroturbulence
parameters, by showing the complexity and variety of line shapes at different locations over
the solar surface (upper panel of Fig.~\ref{Fig:Fe-3D}). 

In the framework of this review, it is of interest to
stress that the predictions of absolute line shifts (due to the unequal surfaces
covered by upward and downward convective motions on the solar surface) are of such good a
quality (Fig.~\ref{Fig:bisector}) that they could even detect erroneous  laboratory wavelengths (by
less than 20~m\AA) for some Fe~I lines \cite{Asplund-00}.

The 3-D simulations of the kind described above \cite{Asplund-00} are of importance not only
for deriving correct abundances in the Sun and late-type stars (see the discussions relative
to iron and oxygen  abundances in Sects.~\ref{Sect:Fe} and \ref{Sect:O}
\cite{Asplund-00a,Bessell02}), but also in the search for extra-solar planets, which
requires to correctly assess the intrinsic radial-velocity jitter.

The methods described above were so far mostly applied to solar-like stars, and not so much
to red giant stars. In these stars where the convective cells are of much larger size
\cite{Schwarzschild-75}, the differences with respect to 1-D models are expected to be even
more dramatic, as the averaging over the stellar surface will not attenuate the effect of
the asymmetries as it does with the solar granulation. Spectacular animations showing these
asymmetries at the surface of a supergiant like Betelgeuse ($\alpha$~Orionis)  were
produced by B.~Freytag ({\tt http://www.astro.uu.se/$\sim$bf/movie/movie.html}). In the
near future, simulations predicting spectral line shapes for giant and supergiant
stars will certainly become available.  We might well face unpleasant surprises when the time comes to compare 
abundances derived in the framework of such 3-D models with the ones obtained with 1-D model atmospheres. 


\section{Available databases of atomic and molecular lines, and
their adequacy to astrophysical applications}
\label{Sect:databases}

There are several papers available in the recent 
literature providing a compilation
of databases of atomic and molecular line lists: 
See \cite{Jorgensen96,VanDishoeck96} for molecules, and
\cite{Cowley03} for atoms.
Other good starting points are the web pages  
{\tt http://cfa-www.harvard.edu/amdata/}\\ {\tt ampdata/amdata.shtml}
from the  {\it Atomic and Molecular Physics Division} of the {\it Harvard Center
for Astrophysics}, and {\tt
  http://plasma-gate.weizmann.ac.il/}\\ {\tt DBfAPP.html} from the {\it Plasma
Laboratory of the Weizmann Institute of Science}, or\\  
{\tt http://www.ster.kuleuven.ac.be/$\sim$leen}
from.  L. Decin's Ph.D. thesis at {\it Leuven University}.

Recent additions 
to this collection of databases include the {\sc
  DREAM} database ({\it Database on Rare Earths At Mons University} 
\cite{Biemont-99,Biemont03}\\
{\tt http://www.umh.ac.be/$\sim$astro/dream.shtml})\\
and Sauval's database (soon to become publicly available on the Royal
Observatory of Belgium web site) providing line identifications 
and solar oscillator strengths in the 1.0--40.0~$\mu$m
range obtained from a fit to the ATMOS solar spectrum (\cite{Farmer89} and
Sect.~\ref{Sect:Sauval}).   
The important impact of these two databases on astrophysical issues has been
discussed in Sects.~\ref{Sect:Li} and \ref{Sect:ISO}.

\begin{figure}
\centerline{\psfig{file=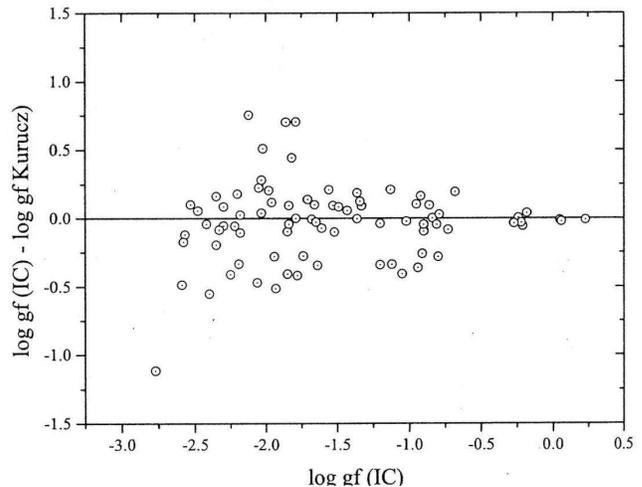,width=8.5cm}}
\caption{\label{Fig:FERRUMKurucz}
Comparison of the laboratory $\log gf$ values for Fe~II lines obtained
by the {\sc FERRUM} project with those from Kurucz.
From \cite{Pickering-01}.
}
\end{figure}

It should be stressed here that not all databases are equally well suited
to the two broad classes of astrophysical applications sketched at the beginning of
Sect.~\ref{Sect:hot}, namely (i)  {\it
opacity calculations}, either in the deep interior or in the atmosphere, and (ii) {\it
spectral synthesis} and {\it abundance determinations}.
Databases providing data which are correct on average for the largest
possible collection of lines (such as the {\sc OPACITY PROJECT/TOPBASE}
\cite{Cowley03,Seaton-94}, {\sc OPAL} \cite{RogersIglesias98}, 
or Kurucz's \cite{Kurucz95}) are necessary
for the former applications, but will generally yield poor results for
the latter, which require accurate  positions {\it and} $f$-values for
specific individual lines. In this case, databases  (like {\sc DREAM},
Sauval's and, to some extent, {\sc VALD} \cite{Kupka-99}) providing
critically-evaluated or measured data  are necessary.
The difficulty here comes from the fact that the lines used to
derive abundances are preferably weak, and are thus
less commonly found among measured data sets than  stronger lines. Bulk
calculations like Kurucz's offer no reliable alternative, since
the quality of $f$ for these weak lines are
generally poorer than for stronger lines. This is well illustrated by Fig.~\ref{Fig:FERRUMKurucz}
which compares Kurucz's oscillator strengths for Fe~II lines with those
measured by the {\sc FERRUM} project \cite{Pickering-01}.  An
illustration of the danger of using $f$-values from bulk computations for
specific line studies has been given in Sect.~\ref{Sect:oldest_stars} in
relation with the V~I line blending the Th~II line used for
cosmochronometry.  The recent breakthrough discussed in
Sect.~\ref{Sect:OMara} offers an interesting way to break this vicious
circle, by making it possible to derive abundances from the {\it wings} of
very strong lines from neutral atoms, in the case that they are
collisionally-broadened by neutral H atoms (Fig.~\ref{Fig:Grevesse}).

\section{Conclusion}

The plea expressed almost 10 years ago as concluding statement of the {\it
Laboratory and Astronomical High Resolution Spectra} conference
\cite{Sauval}, also held in Brussels, remains fully appropriate:
\textsl{We have reached a point where getting accurate 
  laboratory and theoretical data to interpret existing astronomical
  observations is becoming as urgent as getting the astronomical
  spectra themselves}.

\section{Acknowledgments}

I am greatly indebted to E.~Bi\'emont, 
N.~Grevesse, J.~Sauval and S.~Van Eck for very useful 
discussions which greatly enriched the present text. They also
provided me with some unpublished material.  
It is a great pleasure to thank  
M.~Asplund, C.~Cowley, S.~Davis, L.~Decin, and B.~Plez
for helpful comments,  discussions, or for sending useful reprints.



\begin{thebibliography}{100}

\bibitem{Databases-95}
{Adelman}, S.~J. and {Wiese}, W.~L., editors.
\newblock {\em Astrophysical Applications of Powerful New Databases (ASP
  Conf.~Ser.~78)}. ASP: San Francisco, 1995.

\bibitem{Allard97}
{Allard}, F., {Hauschildt}, P.~H., {Alexander}, D.~R., and {Starrfield}, S.
\newblock {\em \araa}, 35, 137, 1997.

\bibitem{Allende01}
{Allende Prieto}, C., {Lambert}, D.~L., and {Asplund}, M.
\newblock {\em \apjl}, 556, L63, 2001.

\bibitem{Anders-Grevesse-89}
{Anders}, E. and {Grevesse}, N.
\newblock {\em \gca}, 53, 197, 1989.

\bibitem{Anstee95}
{Anstee}, S.~D. and {O'Mara}, B.~J.
\newblock {\em \mnras}, 276, 859, 1995.

\bibitem{Aoki2002}
{Aoki}, W., {Ryan}, S.~G., {Norris}, J.~E., {et~al.}
\newblock {\em \apj}, 580, 1149, 2002.

\bibitem{Arnould01}
{Arnould}, M. and {Goriely}, S.
\newblock In {von Hippel}, T., {Simpson}, C., and {Manset}, N., editors, {\em
  Astrophysical Ages and Times Scales (ASP Conf. Ser. 245)}, page 252. ASP: San
  Francisco, 2001.

\bibitem{Asplund-03}
{Asplund}, M. and {Collet}, R.
\newblock In {Cavallo}, R., {Keller}, S., and {Turcotte}, S., editors, {\em 3D
  Stellar Evolution (ASP Conf. Ser.)}, page in press. ASP: San Francisco, 2003.

\bibitem{Asplund-00}
{Asplund}, M., {Nordlund}, {\AA}., {Trampedach}, R., {Allende Prieto}, C., and
  {Stein}, R.~F.
\newblock {\em \aap}, 359, 729, 2000.

\bibitem{Asplund-00a}
{Asplund}, M., {Nordlund}, {\AA}., {Trampedach}, R., and {Stein}, R.~F.
\newblock {\em \aap}, 359, 743, 2000.

\bibitem{Asplund03}
{Asplund}, M., {Sauval}, A.~J., {Grevesse}, N., and {Blomme}, R.
\newblock {\em \aap}, in press, 2003.

\bibitem{Baker-62}
{Baker}, N. and {Kippenhahn}, R.
\newblock {\em Zeitschrift fur Astrophysics}, 54, 114, 1962.

\bibitem{Barklem98b}
{Barklem}, P.~S., {Anstee}, S.~D., and {O'Mara}, B.~J.
\newblock {\em Publ.~Astron.~Soc.~Australia}, 15, 336, 1998.

\bibitem{Barklem97}
{Barklem}, P.~S. and {O'Mara}, B.~J.
\newblock {\em \mnras}, 290, 102, 1997.

\bibitem{Barklem98a}
{Barklem}, P.~S., {O'Mara}, B.~J., and {Ross}, J.~E.
\newblock {\em \mnras}, 296, 1057, 1998.

\bibitem{Bessell02}
{Bessell}, M.~S.
\newblock In {Rickman}, H., editor, {\em Highlights in Astronomy}, volume~12,
  page 410. ASP: San Francisco, 2002.

\bibitem{Biemont-99}
{Bi{\' e}mont}, E., {Palmeri}, P., and {Quinet}, P.
\newblock {\em \apss}, 269, 635, 1999.

\bibitem{Biemont02}
{Bi{\' e}mont}, E., {Palmeri}, P., {Quinet}, P., {Zhang}, Z.~G., and
  {Svanberg}, S.
\newblock {\em \apj}, 567, 1276, 2002.

\bibitem{Biemont-91}
{Bi\'emont}, E., {Baudoux}, M., {Kurucz}, R.~L., {Ansbacher}, W., and
  {Pinnington}, E.~H.
\newblock {\em \aap}, 249, 539, 1991.

\bibitem{Biemont-2000}
{Bi{\'e}mont}, E., {Garnir}, H.~P., {Palmeri}, P., {Li}, Z.~S., and {Svanberg},
  S.
\newblock {\em \mnras}, 312, 116, 2000.

\bibitem{Biemont03}
{Bi{\'e}mont}, E. and {Quinet}, P.
\newblock {\em \physscr}, in press, 2003.

\bibitem{Blackwell84}
{Blackwell}, D.~E., {Booth}, A.~J., and {Petford}, A.~D.
\newblock {\em \aap}, 132, 236, 1984.

\bibitem{Blackwell95a}
{Blackwell}, D.~E., {Lynas-Gray}, A.~E., and {Smith}, G.
\newblock {\em \aap}, 296, 217, 1995.

\bibitem{Blackwell95b}
{Blackwell}, D.~E., {Smith}, G., and {Lynas-Gray}, A.~E.
\newblock {\em \aap}, 303, 575, 1995.

\bibitem{GHRS}
{Brandt}, J.~C., {Ake}, T.~B., and {Peterson}, C.~C., editors.
\newblock {\em The Scientific Impact of the Goddard High-Resolution
  Spectrograph}, ASP Conf. Ser. 143. ASP: San Francisco, 1998.

\bibitem{B2FH}
{Burbidge}, E.~M., {Burbidge}, G.~R., {Fowler}, W.~A., and {Hoyle}, F.
\newblock {\em Rev.~Mod.~Phys.}, 29, 4, 1957.

\bibitem{Butcher87}
{Butcher}, H.~R.
\newblock {\em \nat}, 328, 127, 1987.

\bibitem{Butcher-88}
{Butcher}, H.~R.
\newblock {\em The~Messenger}, 51, 12, 1988.

\bibitem{Cameron-Fowler-71}
{Cameron}, A.~J.~W. and {Fowler}, W.~A.
\newblock {\em \apj}, 164, 111, 1971.

\bibitem{Castro-01}
{Castro-Carrizo}, A., {Bujarrabal}, V., {Fong}, D., {et~al.}
\newblock {\em \aap}, 367, 674, 2001.

\bibitem{Cayrel-01}
{Cayrel}, R., {Hill}, V., {Beers}, T.~C., {et~al.}
\newblock {\em \nat}, 409, 691, 2001.

\bibitem{Cayrel-01a}
{Cayrel}, R., {Spite}, M., {Spite}, F., {et~al.}
\newblock In {von Hippel}, T., {Simpson}, C., and {Manset}, N., editors, {\em
  Astrophysical Ages and Times Scales (ASP Conf.~Ser.~245)}, page 244. ASP: San
  Francisco, 2001.

\bibitem{Christy-66}
{Christy}, R.~F.
\newblock {\em \araa}, 4, 353, 1966.

\bibitem{Cowan96}
{Cowan}, J.~J., {Sneden}, C., {Truran}, J.~W., and {Burris}, D.~L.
\newblock {\em \apjl}, 460, L115, 1996.

\bibitem{Cowley03}
{Cowley}, C.~R., {Adelman}, S.~J., and {Bord}, D.~J.
\newblock In {Piskunov} et~al. \cite{IAUS210}, page in press.

\bibitem{Cox-90}
{Cox}, A.~N. and {Morgan}, S.~M.
\newblock In {Cacciari}, C. and {Clementini}, G., editors, {\em Confrontation
  Between Stellar Pulsation and Evolution (ASP Conf.~Ser.~11)}, page 293. ASP:
  San Francisco, 1990.

\bibitem{Cox-92}
{Cox}, A.~N., {Morgan}, S.~M., {Rogers}, F.~J., and {Iglesias}, C.~A.
\newblock {\em \apj}, 393, 272, 1992.

\bibitem{ISO}
{Cox}, P. and {Kessler}, M.~F., editors.
\newblock {\em The Universe as seen by ISO}, ESA SP-427. ESA: Noordwijk, 1999.

\bibitem{Cunha-03}
{Cunha}, K., {Smith}, V.~V., {Lambert}, D.~L., and {Hinkle}, K.~H.
\newblock {\em \aj}, in press, astro, 2003.

\bibitem{Decin-03}
{Decin}, L., {Vandenbussche}, B., {Waelkens}, C., {et~al.}
\newblock {\em \aap}, 400, 679, 2003.

\bibitem{Decin-00}
{Decin}, L., {Waelkens}, C., {Eriksson}, K., {et~al.}
\newblock {\em \aap}, 364, 137, 2000.

\bibitem{DenHartog02}
{Den Hartog}, E.~A., {Wickliffe}, M.~E., and {Lawler}, J.~E.
\newblock {\em \apjs}, 141, 255, 2002.

\bibitem{UVES}
{D'Odorico}, S., {Cristiani}, S., {Dekker}, H., {et~al.}
\newblock In {Bergeron}, J., editor, {\em Discoveries and Research Prospects
  from 8- to 10-Meter-Class Telescopes (Proc. SPIE Vol. 4005)}, pages 121--130.
  The International Society for Optical Engineering, 2000.

\bibitem{Farmer89}
{Farmer}, C.~B. and {Norton}, R.~H.
\newblock {\em A High-Resolution Atlas of the Sun and the Earth Atmosphere from
  Space. Volume I. The Sun}.
\newblock NASA Reference Publication 1224. NASA Scientific and Technical
  Information Division: Washington D.C., 1989.

\bibitem{Fuhr-88}
{Fuhr}, J.~R., {Martin}, G.~A., and {Wiese}, W.~L.
\newblock {\em J.~Phys.~Chem.~Ref.~Data}, 17, Suppl.~No.~4, 1988.

\bibitem{Garz69}
{Garz}, T., {Holweger}, H., {Kock}, M., and {Richter}, J.
\newblock {\em \aap}, 2, 446, 1969.

\bibitem{Geller92}
{Geller}, M.
\newblock {\em A High-Resolution Atlas of the Sun and the Earth Atmosphere from
  Space. Volume III. Key to the Identification of Solar Features}.
\newblock NASA Reference Publication 1224. NASA Scientific and Technical
  Information Division: Washington D.C., 1992.

\bibitem{Goldschmidt37}
{Goldschmidt}, V.~M.
\newblock {\em Skrifter Norske Videnskaps-Akad. Oslo, Math.-Naturw. Kl.}, 4,
  99, 1937.

\bibitem{Goriely-99a}
{Goriely}, S.
\newblock {\em \aap}, 342, 881, 1999.

\bibitem{Goriely-99b}
{Goriely}, S. and {Clerbaux}, B.
\newblock {\em \aap}, 346, 798, 1999.

\bibitem{Goriely-00}
{Goriely}, S. and {Mowlavi}, N.
\newblock {\em \aap}, 362, 599, 2000.

\bibitem{Grevesse-1998}
{Grevesse}, N. and {Sauval}, A.~J.
\newblock {\em \ssr}, 85, 161, 1998.

\bibitem{Grevesse99}
{Grevesse}, N. and {Sauval}, A.~J.
\newblock {\em \aap}, 347, 348, 1999.

\bibitem{Gustafsson-95}
{Gustafsson}, B.
\newblock In {Adelman} and {Wiese} \cite{Databases-95}, page 347.

\bibitem{Gustafsson98}
{Gustafsson}, B.
\newblock In {Bedding}, T.~R., {Booth}, A.~J., and {Davis}, J., editors, {\em
  Fundamental Stellar Properties (IAU Symp. 189)}, page 261. Kluwer Academic
  Publ.: Dordrecht, 1998.

\bibitem{Gustafsson-75}
{Gustafsson}, B., {Bell}, R.~A., {Eriksson}, K., and {Nordlund}, A.
\newblock {\em \aap}, 42, 407, 1975.

\bibitem{Gustafsson-01}
{Gustafsson}, B. and {Mizuno-Wiedner}, M.
\newblock In {von Hippel}, T., {Simpson}, C., and {Manset}, N., editors, {\em
  Astrophysical Ages and Times Scales (ASP Conf.~Ser.~245)}, page 271. ASP: San
  Francisco, 2001.

\bibitem{Henderson-99}
{Henderson}, M., {Irving}, R.~E., {Matulioniene}, R., {et~al.}
\newblock {\em \apj}, 520, 805, 1999.

\bibitem{Hill-02}
{Hill}, V., {Plez}, B., {Cayrel}, R., {et~al.}
\newblock {\em \aap}, 387, 560, 2002.

\bibitem{Hirata-95}
{Hirata}, R. and {Horaguchi}, T.
\newblock {\em VizieR Online Data Catalog}, VI/69,
  http://cdsweb.u{-}strasbg.fr/htbin/Cat?VI/69, 1995.

\bibitem{Holweger91}
{Holweger}, H., {Bard}, A., {Kock}, M., and {Kock}, A.
\newblock {\em \aap}, 249, 545, 1991.

\bibitem{Holweger95}
{Holweger}, H., {Kock}, M., and {Bard}, A.
\newblock {\em \aap}, 296, 233, 1995.

\bibitem{ASP288}
{Hubeny}, I., {Mihalas}, D., and {Werner}, K., editors.
\newblock {\em Stellar Atmosphere Modeling (ASP Conf.~Ser.~288)}. ASP: San
  Francisco, 2003.

\bibitem{Iglesias-92}
{Iglesias}, C.~A., {Rogers}, F.~J., and {Wilson}, B.~G.
\newblock {\em \apj}, 397, 717, 1992.

\bibitem{Jorgensen96}
{J\o rgensen}, U.~G.
\newblock In {Van Dishoeck}, E.~F., editor, {\em Molecules in Astrophysics:
  Probes and Processes (IAU Symp. 178)}, page 441. Kluwer Academic Publ.:
  Dordrecht, 1996.

\bibitem{Johansson-98}
{Johansson}, S.
\newblock In {Brandt} et~al. \cite{GHRS}, page 155.

\bibitem{Johansson02a}
{Johansson}, S.
\newblock In {Rickman}, H., editor, {\em Highlights in Astronomy, Vol.~12
  (XXIVth General Assembly of the IAU)}, pages 84--87. ASP: San Francisco,
  2002.

\bibitem{Johansson-95}
{Johansson}, S., {Brage}, T., {Leckrone}, D.~S., {Nave}, G., and {Wahlgren},
  G.~M.
\newblock {\em \apj}, 446, 361, 1995.

\bibitem{Johansson02b}
{Johansson}, S., {Derkatch}, A., {Donnelly}, M., {et~al.}
\newblock {\em \physscr}, T100, 71, 2002.

\bibitem{Johansson03}
{Johansson}, S., {Litz{\' e}n}, U., {Lundberg}, H., and {Zhang}, Z.
\newblock {\em \apjl}, 584, L107, 2003.

\bibitem{Johnson01}
{Johnson}, J.~A. and {Bolte}, M.
\newblock {\em \apj}, 554, 888, 2001.

\bibitem{IAUC146}
{J{\o}rgensen}, U.~G., editor.
\newblock {\em Molecules in the Stellar Environment (IAU Colloq.~146)}.
  Springer-Verlag: Berlin (Lecture Notes in Physics vol.~428), 1994.

\bibitem{Jorissen-92}
{Jorissen}, A., {Smith}, V.~V., and {Lambert}, D.~L.
\newblock {\em \aap}, 261, 164, 1992.

\bibitem{Karlsson-01}
{Karlsson}, H., {Sikstrom}, C.~M., {Johansson}, S., {Li}, Z.~S., and
  {Lundberg}, H.
\newblock {\em \aap}, 371, 360, 2001.

\bibitem{Kupka-99}
{Kupka}, F., {Piskunov}, N., {Ryabchikova}, T.~A., {Stempels}, H.~C., and
  {Weiss}, W.~W.
\newblock {\em \aaps}, 138, 119, 1999.

\bibitem{Kurucz89}
{Kurucz}, R.~L.
\newblock In {McNally}, D., editor, {\em Transactions IAU, Vol. XXB}, page 168.
  Kluwer Academic Publ.: Dordrecht, 1989.

\bibitem{Kurucz95}
{Kurucz}, R.~L.
\newblock In {Adelman}, S.~J. and {Wiese}, W.~L., editors, {\em Astrophysical
  Applications of Powerful New Databases}, page 205. ASP Conf. Ser. 78: San
  Francisco, 1995.

\bibitem{Kurucz02a}
{Kurucz}, R.~L.
\newblock {\em \ba}, 11, 101, 2002.

\bibitem{Kurucz02}
{Kurucz}, R.~L.
\newblock In {Schultz}, D.~R., {Krstic}, P.~S., and {Ownby}, F., editors, {\em
  Atomic and Molecular Data and their Applications}, pages 134--143. AIP Conf.
  Proc. 636, 2002.

\bibitem{Lambert78}
{Lambert}, D.~L.
\newblock {\em \mnras}, 182, 249, 1978.

\bibitem{Lambert-94}
{Lambert}, D.~L.
\newblock In {J{\o}rgensen} \cite{IAUC146}, page~1.

\bibitem{Lambert-96}
{Lambert}, D.~L., {Heath}, J.~E., {Lemke}, M., and {Drake}, J.
\newblock {\em \apjs}, 103, 183, 1996.

\bibitem{Lambert-93}
{Lambert}, D.~L., {Smith}, V.~V., and {Heath}, J.
\newblock {\em PASP}, 105, 568, 1993.

\bibitem{Lawler01a}
{Lawler}, J.~E., {Bonvallet}, G., and {Sneden}, C.
\newblock {\em \apj}, 556, 452, 2001.

\bibitem{Lawler90}
{Lawler}, J.~E., {Whaling}, W., and {Grevesse}, N.
\newblock {\em \nat}, 346, 635, 1990.

\bibitem{Lawler01b}
{Lawler}, J.~E., {Wickliffe}, M.~E., {Cowley}, C.~R., and {Sneden}, C.
\newblock {\em \apjs}, 137, 341, 2001.

\bibitem{Lawler01c}
{Lawler}, J.~E., {Wickliffe}, M.~E., {den Hartog}, E.~A., and {Sneden}, C.
\newblock {\em \apj}, 563, 1075, 2001.

\bibitem{Leckrone-93}
{Leckrone}, D.~S., {Johansson}, S.~G., {Wahlgren}, G.~M., and {Adelman}, S.~J.
\newblock {\em \physscr}, T47, 149, 1993.

\bibitem{Leckrone-96}
{Leckrone}, D.~S., {Johansson}, S.~G., {Wahlgren}, G.~M., {Proffitt}, C.~R.,
  and {Brage}, T.
\newblock {\em \physscr}, T65, 110, 1996.

\bibitem{Leckrone-99}
{Leckrone}, D.~S., {Proffitt}, C.~R., {Wahlgren}, G.~M., {Johansson}, S.~G.,
  and {Brage}, T.
\newblock {\em \aj}, 117, 1454, 1999.

\bibitem{Michaud-70}
{Michaud}, G.
\newblock {\em \apj}, 160, 641, 1970.

\bibitem{Morell92}
{Morell}, O., {Kallander}, D., and {Butcher}, H.~R.
\newblock {\em \aap}, 259, 543, 1992.

\bibitem{Moskalik-92}
{Moskalik}, P. and {Dziembowski}, W.~A.
\newblock {\em \aap}, 256, L5, 1992.

\bibitem{Nakajima95}
{Nakajima}, T., {Oppenheimer}, B.~R., {Kulkarni}, S.~R., {et~al.}
\newblock {\em \nat}, 378, 463, 1995.

\bibitem{Nilsson02a}
{Nilsson}, H., {Ivarsson}, S., {Johansson}, S., and {Lundberg}, H.
\newblock {\em \aap}, 381, 1090, 2002.

\bibitem{Nilsson02b}
{Nilsson}, H., {Zhang}, Z.~G., {Lundberg}, H., {Johansson}, S., and {Nordstr{\"
  o}m}, B.
\newblock {\em \aap}, 382, 368, 2002.

\bibitem{Noguchi-02}
{Noguchi}, K., {Aoki}, W., {Kawanomoto}, S., {et~al.}
\newblock {\em \pasj}, 54, 855, 2002.

\bibitem{Nordlund-90}
{Nordlund}, {\AA}. and {Dravins}, D.
\newblock {\em \aap}, 228, 155, 1990.

\bibitem{Norris97b}
{Norris}, J.~E., {Ryan}, S.~G., and {Beers}, T.~C.
\newblock {\em \apjl}, 489, L169, 1997.

\bibitem{Oppenheimer95}
{Oppenheimer}, B.~R., {Kulkarni}, S.~R., {Matthews}, K., and {Nakajima}, T.
\newblock {\em \sci}, 270, 1478, 1995.

\bibitem{Palmeri-00}
{Palmeri}, P., {Quinet}, P., {Fr{\' e}mat}, Y., {Wyart}, J.-F., and {Bi{\'
  e}mont}, E.
\newblock {\em \apjs}, 129, 367, 2000.

\bibitem{Pickering-02}
{Pickering}, J.~C., {Donnelly}, M.~P., {Nilsson}, H., {Hibbert}, A., and
  {Johansson}, S.
\newblock {\em \aap}, 396, 715, 2002.

\bibitem{Pickering-01}
{Pickering}, J.~C., {Johansson}, S., and {Smith}, P.~L.
\newblock {\em \aap}, 377, 361, 2001.

\bibitem{Pickering95}
{Pickering}, J.~C. and {Semeniuk}, J.~I.
\newblock {\em \mnras}, 274, L37, 1995.

\bibitem{Herschel}
{Pilbratt}, G.~L., {Cernicharo}, J., {Heras}, A.~M., {Prusti}, T., and
  {Harris}, R., editors.
\newblock {\em The Promise of the Herschel Space Observatory}, ESA SP-460. ESA:
  Noordwijk, 2001.

\bibitem{IAUS210}
{Piskunov}, N., {Weiss}, W.~W., and {Gray}, D.~F., editors.
\newblock {\em Modeling of Stellar Atmospheres (IAU Symp. 210)}. ASP: San
  Francisco, 2003.

\bibitem{Plez-98}
{Plez}, B.
\newblock {\em \aap}, 337, 495, 1998.

\bibitem{Rebolo95}
{Rebolo}, R., {Zapatero-Osorio}, M.~R., and {Martin}, E.~L.
\newblock {\em \nat}, 377, 129, 1995.

\bibitem{Reyniers02}
{Reyniers}, M., {Van Winckel}, H., {Bi{\' e}mont}, E., and {Quinet}, P.
\newblock {\em \aap}, 395, L35, 2002.

\bibitem{IAU24JD1}
{Rickman}, H., editor.
\newblock {\em Highlights of Astronomy. XXIVth IAU General Assembly, Joint
  Discussion 1. Atomic and Molecular Data for Astrophysics: New Developments,
  Case Studies and Future Needs}. ASP: San Francisco, 2002.

\bibitem{Rogers-Iglesias-92}
{Rogers}, F.~J. and {Iglesias}, C.~A.
\newblock {\em \apj}, 401, 361, 1992.

\bibitem{RogersIglesias98}
{Rogers}, F.~J. and {Iglesias}, C.~A.
\newblock {\em \ssr}, 85, 61, 1998.

\bibitem{Ruiz97}
{Ruiz}, M.~T., {Leggett}, S.~K., and {Allard}, F.
\newblock {\em \apjl}, 491, L107, 1997.

\bibitem{Russell29}
{Russell}, H.~N.
\newblock {\em \apj}, 70, 11, 1929.

\bibitem{Sauval}
{Sauval}, A.~J., {Blomme}, R., and {Grevesse}, N., editors.
\newblock ASP: San Francisco, 1995.

\bibitem{Schramm-74}
{Schramm}, D.~N.
\newblock {\em \araa}, 12, 383, 1974.

\bibitem{Schwarzschild-75}
{Schwarzschild}, M.
\newblock {\em \apj}, 195, 137, 1975.

\bibitem{Seaton-94}
{Seaton}, M.~J., {Yan}, Y., {Mihalas}, D., and {Pradhan}, A.~K.
\newblock {\em \mnras}, 266, 805, 1994.

\bibitem{Smith-96a}
{Smith}, K.~C.
\newblock {\em \apss}, 237, 77, 1996.

\bibitem{Sneden-03}
{Sneden}, C. and {Cowan}, J.~J.
\newblock {\em \sci}, 299, 70, 2003.

\bibitem{Sneden-98}
{Sneden}, C., {Cowan}, J.~J., {Burris}, D.~L., and {Truran}, J.~W.
\newblock {\em \apj}, 496, 235, 1998.

\bibitem{Sneden96}
{Sneden}, C., {McWilliam}, A., {Preston}, G.~W., {et~al.}
\newblock {\em \apj}, 467, 819, 1996.

\bibitem{Sterken-93}
{Sterken}, C. and {Jerzykiewicz}, M.
\newblock {\em Space Science Reviews}, 62, 95, 1993.

\bibitem{Suess-Urey-57}
{Suess}, H.~E. and {Urey}, H.~C.
\newblock {\em Rev.~Mod.~Phys.}, 28, 53, 1956.

\bibitem{Unsold-55}
{Uns{\"o}ld}, A.
\newblock {\em Physik der Sternatmosph{\"a}ren}.
\newblock Springer Verlag: Berlin, 1955.

\bibitem{VanDishoeck96}
{Van Dishoeck}, E.
\newblock In {Van Dishoeck}, E.~F., editor, {\em Molecules in Astrophysics:
  Probes \& Processes (IAU Symp. 178)}, page 539. Kluwer Academic Publ.:
  Dordrecht, 1996.

\bibitem{VanEck-01}
{Van Eck}, S., {Goriely}, S., {Jorissen}, A., and {Plez}, B.
\newblock {\em Nature}, 412, 793, 2001.

\bibitem{VanEck-03}
{Van Eck}, S., {Goriely}, S., {Jorissen}, A., and {Plez}, B.
\newblock {\em \aap}, 404, 291, 2003.

\bibitem{VanHoof-98}
{van Hoof}, P.~A.~M.
\newblock In {Mezzacappa}, A., editor, {\em Stellar Evolution, Stellar
  Explosions and Galactic Chemical Evolution}, page~67. Institute of Physics
  Publishing: Bristol, 1998.

\bibitem{Vogt-94}
{Vogt}, S.~S., {Allen}, S.~L., {Bigelow}, B.~C., {et~al.}
\newblock In {Crawford}, D.~L. and Eric R.~{Craine}, E.~R., editors, {\em Proc.
  SPIE Instrumentation in Astronomy VIII}, volume SPIE Vol. 2198, page 362,
  1994.

\bibitem{Wahlgren-01}
{Wahlgren}, G.~M., {Brage}, T., {Brandt}, J.~C., {et~al.}
\newblock {\em \apj}, 551, 520, 2001.

\bibitem{Wallerstein-97}
{Wallerstein}, G., {Iben}, I., {Parker}, P., {et~al.}
\newblock {\em Rev.~Mod.~Phys.}, 69, 995, 1997.

\bibitem{Westin-00}
{Westin}, J., {Sneden}, C., {Gustafsson}, B., and {Cowan}, J.~J.
\newblock {\em \apj}, 530, 783, 2000.

\bibitem{Wickliffe00}
{Wickliffe}, M.~E., {Lawler}, J.~E., and {Nave}, G.
\newblock {\em \jqsrt}, 66, 363, 2000.

\bibitem{Worrall-73}
{Worrall}, G. and {Wilson}, A.~M.
\newblock {\em Vistas in Astronomy}, 15, 39, 1973.

\end{thebibliography}

\end{document}